\documentclass[journal]{IEEEtran}
\usepackage[cmex10]{amsmath}
\usepackage{longtable,mathdots}
\usepackage{latexsym,enumerate,amssymb,amsbsy}
\usepackage{verbatim,color}
\usepackage{supertabular}
\usepackage{algorithm,algorithmic}
\usepackage{breakurl}
\usepackage{tikz}
\usepackage{url}
\usepackage{yhmath}
\allowdisplaybreaks

\newtheorem{theorem}{Theorem}[section]
\newtheorem{lemma}[theorem]{Lemma}

\newtheorem{definition}[theorem]{Definition}
\newtheorem{remark}[theorem]{Remark}
\newtheorem{proposition}[theorem]{Proposition}

\numberwithin{equation}{section}
\newtheorem{example}[theorem]{Example}

\newcommand{\calX}{\mathcal{X}}
\newcommand{\calW}{\mathcal{W}}
\newcommand{\Z}{\mathbb Z}
\newcommand{\C}{\mathbb C}

\newcommand{\F}{\mathbb F}
\newcommand{\R}{\mathbb R}
\newcommand{\N}{\mathbb N}
\newcommand{\cM}{\mathcal M}
\newcommand{\cE}{\mathcal E}
\newcommand{\Wdual}{{\mathcal{W}^{\perp_{*}}}}
\newcommand{\wt}{\mathop{{\rm wt_{H}}}}

\newcommand{\dimr}{\mathop{{\rm {dim_{\F_{r}}}}}}
\newcommand{\Tr}{\mathop{{\rm Tr}}}
\def\H{\mathop{{_{q/r}{\rm H}}}}
\newcommand{\E}{\mathop{{\rm E}}}
\newcommand{\TrE}{\mathop{{{\rm Tr}_{q/r} \E}}}

\newcommand{\TrEp}{\mathop{{{\rm Tr} \E}}}
\newcommand{\Trq}{\mathop{{{\rm Tr}_{q/r}}}}
\newcommand{\dist}{\mathop{{\rm dist}}}
\renewcommand{\min}{\mathop{{\rm min}}}
\renewcommand{\max}{\mathop{{\rm max}}}
\renewcommand{\u}{\mathbf{u}}

\renewcommand{\v}{\mathbf{v}}
\renewcommand{\c}{\mathbf{c}}
\renewcommand{\d}{\mathbf{d}}

\newcommand{\ab}{{\mathbf{a}}}
\renewcommand{\b}{\mathbf{b}}
\newcommand{\0}{\mathbf{0}}
\newcommand{\1}{\mathbf{1}}
\newcommand{\ddual}{d^{\perp_{*}}}
\newcommand{\bvar}{\boldsymbol{\varphi}}
\newcommand{\bpsi}{\boldsymbol{\psi}}
\newcommand{\CduTrE}{C^{\perp_{\TrE}}}
\newcommand{\bdotcTrE}{\langle \b, \c \rangle_{\TrE}}

\numberwithin{equation}{section}
\title{CSS-like Constructions\\of Asymmetric Quantum Codes}
\author{Martianus~Frederic~Ezerman,~\IEEEmembership{Member,~IEEE}, \linebreak[3]%
        Somphong~Jitman,~\IEEEmembership{}\linebreak[3]%
        San~Ling,~\IEEEmembership{}\linebreak[3]%
        and~Dmitrii~V.~Pasechnik~~\IEEEmembership{}
\thanks{M.~F.~Ezerman was with Laboratoire d'Information Quantique, CP 225, Universit{\'e} 
  Libre de Bruxelles, Av. F. D. Roosevelt 50, B-1050 Belgium. He is now with Centre for Quantum 
  Technologies, National University of Singapore, 3 Science Drive 2, Singapore 117543 
  (email: frederic.ezerman@gmail.com).}%
\thanks{S.~Jitman was with the Division of Mathematical Sciences, School of Physical and Mathematical 
  Sciences, Nanyang Technological University, 21 Nanyang Link, Singapore 637371, Republic of Singapore. 
  He recently joined Department of Mathematics, Faculty of Science, Silpakorn University, Nakhonpathom 73000, 
  Thailand (email: somphong@su.ac.th).}%
\thanks{S.~Ling, and D.~V.~Pasechnik are with the Division of Mathematical Sciences,
  School of Physical and Mathematical Sciences, Nanyang Technological
  University, 21 Nanyang Link, Singapore 637371, Republic of Singapore
  (emails:\{lingsan, dima\}@ntu.edu.sg).}%
\thanks{The work of M.~F.~Ezerman, S.~Jitman and S.~Ling was partially supported
by Singapore National Research Foundation Competitive Research
Program Grant NRF-CRP2-2007-03. D.~V.~Pasechnik's work was supported by
Tier 1 Nanyang Technological University/Singapore Ministry of
Education Grant No. RG60/08. M.~F.~Ezerman and S.~Jitman benefitted from travel 
grants provided through the Merlion Project No.1.02.10 by the Embassy of France in Singapore.}%
\thanks{Copyright (c) 2012 IEEE. Personal use of this material is permitted. 
However, permission to use this material for any other purposes must be obtained from the IEEE 
by sending a request to pubs-permissions@ieee.org.}%
}

\begin{document}

\maketitle

\begin{abstract}
Asymmetric quantum error-correcting codes (AQCs) may offer some advantage over their symmetric 
counterparts by providing better error-correction for the more frequent error types. The well-known 
CSS construction of $q$-ary AQCs is extended by removing the $\F_{q}$-linearity 
requirement as well as the limitation on the type of inner product used. The proposed constructions 
are called \textit{CSS-like constructions} and utilize pairs of nested subfield linear codes 
under one of the Euclidean, trace Euclidean, Hermitian, and trace Hermitian inner products.

After establishing some theoretical foundations, best-performing CSS-like AQCs are constructed. 
Combining some constructions of nested pairs of classical codes and linear programming, many optimal 
and good pure $q$-ary CSS-like codes for $q \in \{ 2,3,4,5,7,8,9\}$ up to reasonable lengths are found. 
In many instances, removing the $\F_{q}$-linearity and using alternative inner products give us pure AQCs 
with improved parameters than relying solely on the standard CSS construction. 
\end{abstract}


\section{Introduction}\label{sec:intro}

Most of the work to date on quantum error-correcting codes (quantum codes) assumes 
that the quantum channel is symmetric, {\it i.e.}, the different types of errors are assumed to 
occur equiprobably. However, recent papers (see~\cite{ESCH07} and~\cite{IM07}, 
for instance) argue that in many qubit systems, phase-flips (or $Z$-errors) occur 
more frequently than bit-flips (or $X$-errors). This leads to the idea of adjusting 
the error-correction to the particular characteristics of the quantum channel and codes 
that take advantage of the asymmetry are called {\it asymmetric quantum codes} (AQCs). 

Steane first hinted at this concept in~\cite{Ste96}. Some results on mostly binary 
AQCs can be found in~\cite{Aly08} and in~\cite{SKR09}. While at the moment there is no 
general agreement on the most appropriate error models for non-qubit asymmetric channels, 
the most established mathematical model in the general qudit systems available is that of 
Wang \textit{et al.}~\cite{WFLX09}.

In the symmetric framework, Steane's seminal work~\cite{Ste96} and that of Calderbank and Shor~\cite{CS96} 
provided the connection between a pair of classical codes and a class of quantum stabilizer codes. 
The construction is now known as the CSS construction which extends naturally 
to the asymmetric case. In~\cite{AA10} Aly and Ashikhmin supply a proof by modifying Steane's 
original proof.

Using a functional approach, a general mathematical characterization and some constructions of 
AQCs from which the CSS construction for AQCs can be derived are given in~\cite{WFLX09}. 
The results have been extended to include constructions from $\F_{r}$-linear
codes over its quadratic extension $\F_{r^{2}}$ in~\cite{ELS10} under the trace 
Hermitian inner product.
   
This present work provides the following contributions:
\begin{enumerate}
 \item We extend the functional approach to include the so-called CSS-like
constructions based on pairs of nested $\F_{r}$-linear codes over $\F_{q}$
where $\F_{r}$ is any subfield of $\F_{q}$. At the same time we relax the condition 
on the inner product used. It is shown that given the appropriate context, the Hermitian, 
trace Hermitian, and trace Euclidean inner products can be utilized as well.
 \item The extensions lead to pure AQCs with better parameters than relying solely on 
the best ones obtainable from the standard CSS construction. This justifies the effort of 
considering $\F_{r}$-linear pairs of nested codes over $\F_{q}$ and their duals under various inner products.
 \item Of purely mathematical interest, our investigation leads to a better structural 
understanding of the functional approach to AQCs. A diagram detailing the relationships among 
different CSS-like constructions is given in Section~\ref{sec:part_one}.
\item Lists of good pure CSS-like AQCs up to some computationally reasonable lengths 
for $q \in \{2,3,4,5,7,8,9\}$ are given.
\end{enumerate}

The paper is organized into seven sections and four appendices. After this introductory section, 
some preliminary notions from classical coding theory and some basics on the AQC error model are given in 
Section~\ref{sec:prelims}. Section~\ref{sec:part_one} accomplishes several important tasks. First, a brief 
review of both the standard CSS construction and the functional characterization of AQCs is supplied for 
convenience. The CSS-like constructions are then proved and their interconnections are shown. 

Three systematic constructions of nested pairs of  linear and  subfield linear codes are presented in 
Section~\ref{sec:part_three} as main ingredients for the CSS-like constructions.  
A linear programming bound as a measure of the optimality of AQCs is derived in Section~\ref{sec:LP}. 
Combining the results of these two sections, good pure CSS-like codes are listed explicitly with their 
corresponding pair of nested classical codes in Section~\ref{sec:Optimal}. The last section contains 
some conclusion and open problems. The appendices establish results needed in the paper whose detailed 
justifications may distract us from the paper's main lines of thought.

\section{Preliminaries}\label{sec:prelims}

Throughout this work, let $p$ be a prime number and let $\F_{p} \subseteq \F_{r} \subseteq \F_{q}$ with 
$r=p^{l}$ and $q=r^{m}$ be finite fields. The trace mapping $\Trq: \F_{q} \to \F_{r}$ is given by 
$\Trq (\beta) = \beta + \beta^{r}+\beta^{r^{2}}+\ldots+\beta^{r^{m-1}}$.
The subscript $q/r$ is omitted whenever $r=p$ and $q$ is clear.
Important properties of the trace mapping can be found in~\cite[Th. 2.23]{LN97}.

If $q=r^2$, let $\overline{a}$ denote $a^r$ for all $a \in \F_{q}$.
For $\u=(u_{1},u_{2},\ldots,u_{n}) \in \F_{q}^{n}$, $\overline{\u}$
stands for $(\overline{u_1},\overline{u_2},\ldots,\overline{u_n})$. Hence, for
any nonempty set $C \subseteq \F_{q}^n$, $\overline{C}:=\{ \overline{\c} : \c \in C \}$.

\subsection{Coding Theory}\label{subsec:coding}
Given $\u,\v \in \F_{q}^{n}$, let $\wt(\v)$ denote the \textit{Hamming weight} of $\v$ 
and $\dist(\u,\v)$ denote their \textit{Hamming distance}. A code $C$ of length $n$ over 
$\F_{q}$ is a nonempty subset of $\F_{q}^{n}$. The \textit{minimum distance} $d(C)$ 
is given by
\begin{equation*}
d(C) = \min \{ \dist(\u,\v): \u,\v \in C,\u \neq \v \}\text{.}
\end{equation*}
For two distinct codes $C$ and $D$, $\wt(C \setminus D)$ denotes $\min \{ \wt(\u) :
\u\in C \setminus D, \u \neq \0\}$.

An $[n,k,d]_q$-linear code $C$ is a $k$-dimensional $\F_{q}$-subspace of $\F_{q}^n$ with minimum distance $d$. 
For a general, not necessarily $\F_{q}$-linear, code $C \subseteq \F_{q}^{n}$, the notation
$(n,M=|C|,d)_q$ is commonly used. A code $C$ is an \textit{$\F_{r}$-linear
code over $\F_{q}$} if $C$ is a subspace of the $\F_{r}$-vector space $\F_{q}^{n}$. 
When $r$ is clear from the context, $C$ is said to be a \textit{subfield linear code} over $\F_{q}$.

For $\u=(u_1,u_2,\ldots,u_n)$ and $\v=(v_1,v_2,\ldots,v_n)$ in $\F_{q}^n$,
we define the following inner products:
\begin{enumerate}
 \item $\langle \u, \v \rangle_{\E}:=\sum_{i=1}^{n} u_i v_i$ is
the \textit{Euclidean inner product} of $\u$ and $\v$.
 \item $\langle \u, \v \rangle_{\TrE}:=\Trq \left( \langle \u,
\v \rangle_{\E} \right)$ is the \textit{trace Euclidean inner product}
of $\u$ and $\v$ valued in $\F_{r}$.
 \item When $\F_{q}$ is a quadratic extension of $\F_{r}$,
$\langle \u, \v \rangle_{\rm H}:= \sum_{i=1}^{n} u_i \overline{v_i}
=\langle \u, \overline{\v} \rangle_{\E}$ is
the \textit{Hermitian inner product} of $\u$ and $\v$.
 \item Let $q=r^{2}$. Then there are two cases of \textit{trace Hermitian inner product}
depending on the field characteristic:
	\begin{enumerate}
 		\item For even $q$, $\langle \u, \v \rangle_{\Tr \H}:=\Trq (\langle \u, {\v}\rangle_{\rm H})$.
 		\item For odd $q$, $\langle \u, \v \rangle_{\Tr \H}:=\Trq(\alpha \langle \u, {\v}\rangle_{\rm H})$ where $\alpha\in\F_{q} \setminus \{0\}$ is such that $\overline{\alpha}=-\alpha$.
	\end{enumerate}
\end{enumerate}

Let $C \subseteq \F_{q}^{n}$ be a code. Let $*$ represent one of the Euclidean, trace Euclidean,
Hermitian and trace Hermitian inner products, the \textit{dual code}
$C^{\perp_{*}}$ of $C$ is given by
\begin{equation*}
 C^{\perp_{*}} := \left\lbrace \u \in \F_q^n : \left\langle
\u,\v\right\rangle _{*} = 0 \text{ for all } \v \in C
\right\rbrace
\end{equation*}
while the dual distance $d^{\perp_{*}}$ is defined to be $d(C^{\perp_{*}})$.

If $C \subseteq C^{\perp_{*}}$, then $C$ is said to be \textit{self-orthogonal}.
$C$ is \textit{self-dual} when equality holds.

A code $C$ is \textit{closed} under $*$ if $(C^{\perp_{*}})^{\perp_{*}} = C$. 
The closure property of linear codes under the Euclidean and Hermitian inner products 
and of subfield linear codes under the trace Hermitian is well known from~\cite[Ch. 3]{NRS06}. 
The said property of subfield linear codes under the trace Euclidean inner product will 
be established in Theorem~\ref{thm:closure}.

\begin{definition}\label{def2.2}
The \textit{weight enumerator} $W_C(X,Y)$ of an $(n,M,d)_q$-code
$C$ is the polynomial
\begin{equation}\label{WE}
W_C(X,Y)=\sum_{i=0}^n A_{i} X^{n-i}Y^{i} \text{,}
\end{equation}
where $A_{i}:=|\{\c \in C : \wt(\c)=i\}|$.
\end{definition}

\begin{theorem}\label{thm:closure}
Let $C$ be an $\F_{r}$-linear code over $\F_{q}$. Then, under the trace Euclidean inner product,
\begin{equation}\label{eq:MacWTrE}
W_{\CduTrE} (X,Y)= \frac{1}{|C|} W_C(X+(q-1)Y,X-Y) \text{.}
\end{equation}
Moreover, $(\CduTrE)^{\perp_{\TrE}} = C$.
\end{theorem}
\begin{IEEEproof}
The proof can be found in Appendix A.
\end{IEEEproof}

In light of Theorem~\ref{thm:closure}, under $*$, the weight enumerator of the dual code of 
a linear or subfield linear $(n,M=|C|,d)_{q}$-code $C$ is connected to the weight 
enumerator of the code $C$ via the MacWilliams Equation
\begin{equation}\label{eq:MacW}
 W_{C^{\perp_{*}}} (X,Y)= \frac{1}{|C|} W_C(X+(q-1)Y,X-Y) \text{.}
\end{equation}

For $0 \leq j \leq n$, let $A_{j}^{\perp_{*}}$ denote the number of codewords
of weight $j$ in $C^{\perp_{*}}$. Then
\begin{equation}\label{eq:Krawteq}
A_{j}^{\perp_{*}} = \frac{1}{|C|} \sum_{i=0}^{n} A_{i} K_{j}^{n,q}(i)
\end{equation}
where $K_{j}^{n,q}(i)$, the {\it Krawtchouk polynomial} of degree $j$ in variable $i$, is given by
\begin{equation}\label{eq:Krawt}
K_{j}^{n,q}(i):= \sum_{l=0}^{j} (-1)^{l} (q-1)^{j-l} \binom{i}{l} \binom{n-i}{j-l}\text{.}
\end{equation}

The last two equations will feature prominently in the linear programming set-up in Section~\ref{sec:Optimal}.

\subsection{Asymmetric Quantum Codes}\label{subsec:ASQC}

Let $\C$ be the field of complex numbers and
$\eta=e^{\frac{2\pi \sqrt{-1}}{p}}\in \C$.
We fix an orthonormal basis of $\C^{q}$ 
$$\left\{|\varphi\rangle:\varphi\in \F_{q}\right\}$$
with respect to the Hermitian inner product on $\C^{q}$.
For $n \in \N$, let $V_{n}=(\C^{q})^{\otimes n}$ 
be the $n$ fold tensor product of $\C^{q}$.
Then we can choose the following orthonormal basis for $V_{n}$
\begin{equation*}\label{basis}
\left\{|\c\rangle =|c_{1} c_{2}\ldots c_{n}\rangle
: \c=(c_{1},\ldots,c_{n}) \in \F_{q}^n\right\} \text{,}
\end{equation*}
where $|c_{1} c_{2}\ldots c_{n}\rangle$ abbreviates
$|c_{1}\rangle\otimes|c_{2}\rangle\otimes\cdots \otimes
|c_{n}\rangle$.

For two quantum states $|\bvar\rangle$ and $|\bpsi\rangle$ in 
$V_{n}$ with 
\begin{equation*}
|\bvar\rangle=\sum\limits_{\c\in \F_{q}^{n}}\alpha(\c)|\c\rangle 
\text { and } |\bpsi\rangle=\sum\limits_{\c\in \F_{q}^{n}}\beta(\c)|\c\rangle \text{,}
\end{equation*}
where $\alpha(\c),\beta(\c)\in \C$, the Hermitian inner product of 
$|\bvar\rangle$ and $|\bpsi\rangle$ is given by
\begin{equation*}
\langle \bvar|\bpsi\rangle=\sum\limits_{\c\in \F_{q}^{n}}
\widetilde{\alpha(\c)}\beta(\c)\in \C \text{,} 
\end{equation*}
where  $\widetilde{\alpha(\c)}$ is the complex conjugate of 
$\alpha(\c)$. We say $|\bvar\rangle$ and $|\bpsi\rangle$ are 
\textit{orthogonal} if $\langle \bvar|\bpsi\rangle=0$.

To measure the performance of a quantum code, an appropriate error model
must be chosen (see~\cite{WFLX09} for instance). In defining an AQC $Q$, one considers the
set of error operators that $Q$ can handle. First, a good basis
$\cE_{n}$ of the vector space of complex $q^{n} \times q^{n}$ matrices
$\cM_{q^{n}} (\C)$ needs to be chosen.  Let $a, b \in \F_{q}$. The
unitary operators $X(a)$ and $Z(b)$ on $\C^{q}$ are defined by
\begin{equation}\label{eq:ErOp}
X(a) |\varphi \rangle = |\varphi+a \rangle \text{ and } Z(b)|\varphi\rangle =
\eta^{\left(\langle b,\varphi \rangle_{\TrEp} \right)} |\varphi\rangle \text{.}
\end{equation}
Based on (\ref{eq:ErOp}), for $\ab=(a_1,\ldots,a_{n}) \in \F_{q}^{n}$,
we can write $X(\ab) = X(a_{1}) \otimes \ldots \otimes X(a_{n})$ and
$Z(\ab) = Z(a_{1}) \otimes \ldots \otimes Z(a_{n})$ for the tensor product of
$n$ error operators. The set $\cE_{n}:=\{ X(\ab)Z(\b) : \ab,\b \in \F_{q}^{n} \}$
can be taken as a good error basis.

The error group $G_{n}$ of order $pq^{2n}$ is generated by the matrices in $\cE_{n}$
\begin{equation*}
 G_{n}:=\{ \eta^{c} X(\ab) Z(\b) : \ab,\b \in \F_{q}^{n}, c \in \F_{p} \} \text{.}
\end{equation*}
Let $E = \eta^{c} X(\ab) Z(\b) \in G_{n}$. Then the \textit{quantum weight}
${\rm wt_{Q}}(E)$ of $E$ is given by $|\{ 1 \leq i \leq n : (a_{i},b_{i}) \neq (0,0) \}|$. 
The number of $X$-errors ${\rm wt_{X}}(E)$ and the number of $Z$-errors ${\rm wt_{Z}}(E)$ 
in the error operator $E$ are given, respectively, by $\wt (\ab)$ and $\wt (\b)$. A formal 
definition of $q$-ary AQC can now be given.

\begin{definition}\label{def:AQC}
A \textit{$q$-ary  quantum code} of length $n$ is a subspace $Q$ of
$V_{n}$ with dimension $K \geq1$. Let $d_{x}$ and $d_{z}$ be positive integers. A
quantum code $Q$ in $V_{n}$ is called an 
\textit{asymmetric quantum code} with parameters $((n,K,d_{z}/d_{x}))_{q}$
or $[[n,k,d_{z}/d_{x}]]_{q}$, where $k=\log_{q}K$, if $Q$ detects $d_{x}-1$
qudits of $X$-errors and, at the same time, $d_{z}-1$
qudits of $Z$-errors, \textit{i.e.}, if $\langle \bvar|\bpsi \rangle=0$ for
$|\bvar\rangle,|\bpsi\rangle\in Q$, then $|\bvar\rangle$ and $E|\bpsi\rangle$ 
are orthogonal for any $E \in G_{n}$ such that ${\rm wt_{X}}(E)\leq d_{x}-1$ 
and ${\rm wt_{Z}}(E)\leq d_{z}-1$. 
Such an asymmetric quantum code $Q$ with dimension $K\geq2$ is called \textit{pure} if $|\bvar\rangle$ 
and $E|\bpsi\rangle$ are orthogonal for any $|\bvar\rangle,|\bpsi\rangle\in Q$ and
any $E \in G_{n}$ such that ${\rm wt_{Q}}(E) \geq 1$ and $E$ satisfies
\begin{equation*}
\begin{cases}
{\rm wt_{X}}(E)\leq d_{x}-1 \\ 
{\rm wt_{Z}}(E)\leq d_{z}-1 
\end{cases}\text{.}
\end{equation*}
By convention, an asymmetric quantum code $Q$ with $K=1$ is assumed to be pure.
\end{definition}

\section{CSS-like Constructions}\label{sec:part_one}

This section constitutes the most technical part of the paper. Note that a main tool in 
the derivation of the standard CSS construction from the functional approach in~\cite{WFLX09} 
is the connection between codes and orthogonal arrays (OAs) due to Delsarte (see~\cite[Th. 4.5]{Del73} 
or~\cite[Th. 4.9]{HSS99}). The codewords in a general code $C$ can be seen as the rows of an 
OA $\mathcal{A}$ and vice versa. Since in the construction of the OA $\F_{q}$-linearity 
is not strictly required and the duality can be defined over any valid bilinear form, it is of 
mathematical interest to investigate if the CSS construction can be extended by relaxing the 
linearity requirement and including other types of inner products.

First, we derive a construction of pure AQCs based on nested pairs of codes over $\F_{q}$ 
under the trace Euclidean inner product. Then, we show how this construction is related to other known 
extensions of the CSS construction discussed in~\cite{WFLX09} and in~\cite{ELS10}.

Recall the following characterization of AQCs presented in~\cite{WFLX09}.
\begin{theorem}\cite[Th. 3.1]{WFLX09}\label{thm:prelim1}
\begin{enumerate}
 \item There exists an asymmetric quantum code with parameters
$((n,K,d_z/d_x))_q$ with $K \geq 2$ if and only if there exist
$K$ nonzero mappings
\begin{equation}\label{eq:3.1}
\varphi_i : \F_q^n \rightarrow \mathbb{C} \text{ for } 1\leq i \leq K
\end{equation}

satisfying the following conditions: for each $d$ such that
$1 \leq d \leq \min\left\lbrace d_x,d_z\right\rbrace $ and
partition of $\left\lbrace 1,2,\ldots,n\right\rbrace $,
\begin{equation}\label{eq:3.2}
\begin{cases}
\left\lbrace 1,2,\ldots,n \right\rbrace = A \cup X \cup Z \cup B \\
|A| = d-1,\quad |B| = n+d-d_x-d_z+1 \\
|X|=d_x - d,\quad |Z| = d_z - d
\end{cases}\text{,}
\end{equation}

and each $\c_A,\c_A' \in \F_q^{|A|}$,
$\c_Z \in \F_q^{|Z|}$ and $\ab_X \in \F_q^{|X|}$,
we have the equality
\begin{multline}\label{eq:3.3}
\sum_{\substack{ \c_X \in \F_q^{|X|} \text{,}\\ \c_B \in \F_q^{|B|}}}
\widetilde{\varphi_i (\c_A,\c_X,\c_Z,\c_B)} \varphi_j(\c_A',\c_X - \ab_X,\c_Z,\c_B) \\
=
\begin{cases}
0 &\text{for $i \neq j$} \\
I(\c_A,\c_A',\c_Z,\ab_X) &\text{for $i = j$}
\end{cases}\text{,}
\end{multline}
where $I(\c_A,\c_A',\c_Z,\ab_X)$ is an element of $\mathbb{C}$
which is independent of $i$. The notation $(\c_A,\c_X,\c_Z,\c_B)$
represents the rearrangement of the entries of the
vector $\c \in \F_q^n$ according to the partition of
$\left\lbrace 1,2,\ldots,n\right\rbrace $ given in
(\ref{eq:3.2}).

\item Let $(\varphi_i,\varphi_j)$ stand for
$\sum_{\c\in \F_q^n} \widetilde{\varphi_i(\c)}\varphi_j(\c)$.
There exists a pure asymmetric quantum code with
parameters $((n,K\geq1,d_z/d_x))_q$
if and only if there exist $K$ nonzero mappings
$\varphi_i$ as shown in (\ref{eq:3.1}) such that
\begin{itemize}
 \item $\varphi_i$ are linearly independent for
$1\leq i\leq K$, i.e., the rank of the $K \times q^n$ matrix
 $(\varphi_i (\c))_{1\leq i\leq K, \c \in \F_q^n}$ is $K$;
and
 \item for each $d$ with
$1 \leq d \leq \min\left\lbrace d_x,d_z\right\rbrace $,
a partition in (\ref{eq:3.2})
and $\c_A,\ab_A \in \F_q^{|A|}, \c_Z \in \F_q^{|Z|}$
and $\ab_X \in \F_q^{|X|}$, we have the equality
\end{itemize}
\end{enumerate}
\begin{multline}\label{eq:3.4}
\sum_{\substack{ \c_X \in \F_q^{|X|}, \\ \c_B \in
\F_q^{|B|}}} \widetilde{\varphi_i (\c_A,\c_X,\c_Z,\c_B)}
\varphi_j(\c_A+\ab_A,\c_X + \ab_X,\c_Z,\c_B) \\
=
\begin{cases}
0 &\text{for $(\ab_A,\ab_X) \neq (\0,\0)$} \\
\frac{(\varphi_i,\varphi_j)}{q^{d_z-1}}
&\text{for $(\ab_A,\ab_X) = (\0,\0)$}
\end{cases}\text{.}
\end{multline}
\end{theorem}

\begin{remark}\label{rem:int}
It is important to note that the values $d_{x}$ and $d_{z}$ are 
in fact interchangeable~\cite[Prop. 4.2]{ELS10}. Physically, such an 
interchange can be effected by applying the Hadamard transform. 
In the presentation of the parameters of a particular AQC, it is customary 
to write $d_{z} \geq d_{x}$ since phase-flip errors are taken to be more frequent. 
\end{remark}

\begin{theorem}\label{thm:prelim2}
Let $d_x,d_z \in \N$. Let $C$ be
an $\F_{r}$-linear code over $\F_q$ of length $n$. Assume
that $d^{\perp_{\TrE}}= d(C^{\perp_{\TrE}})$ is the minimum
distance of the dual code $C^{\perp_{\TrE}}$ of $C$
under the trace Euclidean inner product. For a set
$V:=\left\lbrace \v_i : 1\leq i \leq K\right\rbrace $
of $K$ distinct vectors in $\F_q^n$, let
$d_v:=\min\left\lbrace \wt (\v_i - \v_j + \c)
: 1 \leq i \neq j \leq K, \c \in C\right\rbrace$.
If $d^{\perp_{\TrE}} \geq d_z$ and $d_v \geq d_x$,
then there exists an asymmetric quantum code $Q$ with
parameters $((n,K,d_z/d_x))_q$.
\end{theorem}

\begin{IEEEproof}
The proof follows the same line of argument as the proof of~\cite[Th. 4.4]{ELS10}, 
substituting the trace Euclidean inner product for the trace Hermitian inner product. 
The key reason why the same argument works lies in the usage of the close connection 
between codes and orthogonal arrays~\cite[Th. 4.9]{HSS99} under any valid bilinear form. 
Furthermore, the said connection guarantees that the conditions in Part 2) of 
Theorem~\ref{thm:prelim1} are satisfied, making the resulting AQCs pure.
\end{IEEEproof}

\begin{theorem}\label{thm:AQCTrEu}
For $i=1,2$, let $C_{i}$ be an $\F_{r}$-linear code with parameters $(n,K_{i},d_{i})_{q}$.
If $C_{1}^{\perp_{\TrE}} \subseteq C_{2}$, then there exists an asymmetric quantum code $Q$ with
parameters $\left(\left( n, \frac{K_{1} \cdot K_{2}}{q^{n}}, d_{2}/d_{1}\right)\right)_{q} =
[[n, \log_{q}K_{1}+ \log_{q}K_{2}-n,d_{2}/d_{1}]]_{q}$.
\end{theorem}

\begin{IEEEproof}
We take $C = C_1^{\perp_{\TrE}}$ in Theorem~\ref{thm:prelim2} above.
Since $C_1^{\perp_{\TrE}} \subseteq C_2$, we have
$C_2 = C_1^{\perp_{\TrE}} \oplus C'$, where $C'$ is an $\F_{r}$-subspace
of $C_2$ and $\oplus$ is the direct sum so that
$|C'|=\frac{|C_2|}{\left| C_1^{\perp_{\TrE}} \right|}$.
Let $C' = \left\lbrace \v_1,\ldots,\v_K\right\rbrace$,
where $K = \frac{|C_2|}{\left|C_1^{\perp_{\TrE}}\right|}= \frac{K_{1} \cdot K_{2}}{q^{n}}$
by Theorem~\ref{thm:closure}. Then
\begin{equation*}
d^{\perp_{\TrE}} = d(C^{\perp_{\TrE}}) = d(C_1) = d_1 \text{ and}
\end{equation*}
\begin{align*}
d_v &= \min\left\lbrace \wt (\v_i - \v_j + \c) :
1 \leq i \neq j \leq K, \c \in C \right\rbrace \\
    &= \min \left\lbrace \wt (\v + \c) :
\0 \neq \v \in C', \c \in C_1^{\perp_{\TrE}} \right\rbrace \geq d_2 \text{.}
\end{align*}
\end{IEEEproof}

The standard CSS construction for pure asymmetric $q$-ary quantum codes 
employs the pair $C_{1}^{\perp_{\E}} \subseteq C_{2}$ of $\F_{q}$-linear codes of length $n$.

\begin{theorem}(Standard CSS Construction for AQC)\label{thm:standardCSS}
Let $C_{i}$ be $\F_{q}$-linear codes with parameters $[n,k_{i},d_{i}]_{q}$ for $i=1,2$ 
with $C_{1}^{\perp_{\E}}\subseteq C_{2}$. Let 
\begin{equation*}\label{eq:distances}
d_{z}:= \wt(C_{2} \setminus C_{1}^{\perp_{\E}}) \text{ and } d_{x}:= \wt(C_{1} \setminus C_{2}^{\perp_{\E}})\text{.}
\end{equation*}	
Then there exists an AQC $Q$ with parameters $[[n,k_{1}+k_{2}-n,d_{z}/d_{x}]]_{q}$. The code $Q$ is 
\textit{pure} whenever $d_{z} = d_{2}$ and $d_{x} = d_{1}$.
\end{theorem}

A proof for this construction for the pure case using the functional approach is given in~\cite[Cor. 3.3]{WFLX09}. 
When $q=r^{2}$, we can use either the Euclidean or the Hermitian inner product in the statement of 
Theorem~\ref{thm:standardCSS}. 

Noting that the trace Euclidean inner product is just the Euclidean inner 
product when the codes involved are $\F_{q}$-linear,~\cite[Cor. 3.3]{WFLX09} 
follows immediately from Theorem~\ref{thm:AQCTrEu}. 

Let $\F_{q}$ be a quadratic extension of $\F_{r}$. Let the codes in the 
nested pair be $\F_{r}$-linear codes in $\F_q^{n}$. Under the trace Hermitian 
inner product, we can derive AQCs according to~\cite[Th. 4.5]{ELS10}. When 
the codes are $\F_q$-linear, the trace Hermitian duals become the Hermitian duals. 
Hence, the construction with respect to the Hermitian inner product follows.

In summary, the standard CSS construction for pure AQCs can be extended to 
include the constructions of pure AQCs from nested pairs of classical codes 
under the Hermitian, trace Hermitian, and trace Euclidean inner products. 
We call all of the above constructions \textit{CSS-like}.

To show the generality of Theorem~\ref{thm:AQCTrEu}, we demonstrate how to 
derive~\cite[Th. 4.5]{ELS10} when $q=r^{2}$ from it. Given a nested pair 
$C_{1}^{\perp_{\TrE}} \subseteq C_{2}$ of codes yielding a quantum code of 
parameters $((n,K,d_{z}/d_{x}))_{q}$ we construct a nested pair 
$D_{1}^{\perp_{\Tr \H}} \subseteq D_{2}$ of codes yielding a quantum code 
of equal parameters and vice versa.

\begin{theorem}\label{thm:TrEandTrH}
Let $q=r^{2}$. Then an $((n,K,d_{z}/d_{x}))_{q}$-CSS-like quantum code
with respect to the trace Euclidean inner product exists
if and only if there exists an $((n,K,d_{z}/d_{x}))_{q}$-CSS-like quantum code
with respect to the trace Hermitian inner product.
\end{theorem}

\begin{IEEEproof}
See Appendix B.
\end{IEEEproof}

If, in Theorem~\ref{thm:TrEandTrH}, the codes in the nested pairs are
$\F_{q}$-linear, then we get the link between AQCs based on the CSS-like 
constructions under the Hermitian and Euclidean inner products.

The mathematical structures investigated above reveal that for a pair of 
nested $\F_{q}$-linear codes it suffices to consider the Euclidean inner product. 
In all other cases, it suffices to use the trace Euclidean inner product.
 
\begin{figure}[h!]
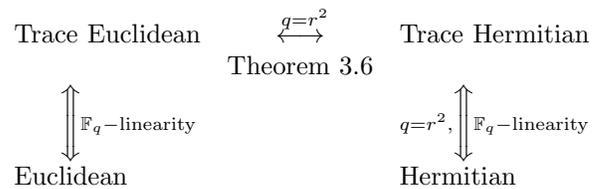

\begin{center}
\begin{align*}
\begin{array}[c]{lcl}
{\rm Trace~Euclidean}                       &\stackrel{~q=r^2}{\longleftrightarrow}     &{\rm Trace~Hermitian}\\
                                            &{\rm Theorem~\ref{thm:TrEandTrH}}          &\\
{~~~~~}\Bigg\Updownarrow\scriptstyle{\F_{q}-\rm linearity}&                             &\scriptstyle{q=r^2,}\Bigg\Updownarrow\scriptstyle{\F_{q}-\rm linearity}\\
{\rm Euclidean}                             &                                           & {\rm Hermitian}
\end{array}
\end{align*}
\end{center}
\caption{Relationships among CSS-like Constructions}
\label{fig:one}
\end{figure}

The relationships among different CSS-like constructions is summarized in Fig.~\ref{fig:one} 
with the horizontal arrow signifying that the resulting AQCs have the same parameters.

Applying a suitable CSS-like construction to the pair $C \subseteq C^{\perp_{*}}$
gives us the following proposition.
\begin{proposition}\label{prop:Sorth}
Let $C$ be a self-orthogonal $(n,|C|,d)_{q}$-code. Then there exists
an AQC $Q$ with parameters $[[n,n-2\log_{q}(|C|),d^{\perp_{*}}/d^{\perp_{*}}]]_{q}$.
\end{proposition}

The existence of some pure CSS-like AQCs with specified parameters can often 
be ruled out by examining the parameters of the component codes in the nested pair used. 
\begin{example}
There does not exist a pure $[[5,1,3/2]]_{2}$-CSS code.
\end{example}
\begin{IEEEproof}
First, note that there is no codeword $\v$ of weight $5$ in any $[5,2,3]_{2}$-code $C$. 
Let $\c \in C$ such that $\wt(\c)=3$. If such a codeword $\v$ exists, then $\c + \v$ is 
a codeword of weight $2$ in $C$, a contradiction. Another possibility is a nested pair 
$[5,1,d]_{2} \subset [5,2,3]_{2}$ with $d^{\perp_{\E}}=2$. But this forces $d=5$, 
which has been shown to be impossible above. Since $k \leq n-d+1$ by the Singleton bound, the 
remaining candidates of nested pairs, namely  $[5,1,d]_{2} \subset [5,2,2]_{2}$ with $d^{\perp_{\E}}=3$, 
$[5,2,d]_{2} \subset [5,3,2]_{2}$ with $d^{\perp_{\E}}=3$, and $[5,3,d]_{2} \subset [5,4,2]_{2}$ 
with $d^{\perp_{\E}}=3$, can all be shown to be impossible.
\end{IEEEproof}

The next example provides a partial answer to a question raised in~\cite[p. 1652]{SKR09}.
\begin{example}
A pure $[[12,1,5/3]]_2$-CSS code does not exist.
\end{example}
\begin{IEEEproof}
For a contradiction, assume that such a code exists. Then we have
a pair of binary classical codes $C_1$ with parameters
$[12,k_{1},d_{1}]_{2}$ and $C_2$ with parameters $[12,k_{2},d_{2}]_{2}$,
such that $C_{1}^{\perp_{\E}} \subset C_{2}$ with $k_{1} + k_{2}-12 = 1$ and
$\{d_{1},d_{2}\}=\{3,5\}$.

Case 1: $d_1=3$ and $d_2=5$:
From~\cite{Gr09}, $2 \leq k_2 \leq 4$. This forces $9 \leq k_1 \leq 11$.
However, for $[12,k_{1},d_{1}]_{2}$ with $9 \leq k_1 \leq 11$, $d_1 \leq 2 < 3$.

Case 2: $d_1=5$ and $d_2=3$:
From~\cite{Gr09}, $2 \leq k_2 \leq 8$. This forces $5 \leq k_1 \leq 11$.
However, for $[12,k_{1},d_{1}]_{2}$ with $5 \leq k_1 \leq 11$, $d_1 \leq 4 < 5$.
\end{IEEEproof}

Now that the theoretical foundations on the CSS-like constructions have been established, 
we next show that there are indeed gains on the parameters of the resulting AQCs. A two-directional 
approach is employed in coming up with such AQCs. First, we directly construct nested pairs of 
classical codes and derive the parameters of the resulting AQCs in the next section. Linear programming 
is then used to derive the upper bound for $\log_{q}(K)$ in the section after next.

\section{Three Constructions of Nested Pairs of Codes}\label{sec:part_three}
In this section, we derive pairs of linear and subfield linear codes which can be used to 
construct AQCs. Three constructions are considered, namely a construction based on nested 
cyclic $\F_{r}$-linear codes over $\F_{q}$, a construction from nested group 
character codes, and a construction based on best-known linear codes (BKLC) of 
length $n$ having a codeword $\v$ such that $\wt(\v)=n$. This last construction 
yields AQCs with $d_{x}=2$. All computations are done in MAGMA~\cite{BCP97} version V2.16-5.

\subsection{Cyclic Construction}\label{subsec:cyclic}
An obvious choice for the construction of nested pairs of $\F_{q}$-linear 
codes is the cyclic construction. Earlier construction of AQCs 
based on $\F_{2}$-cyclic codes has been done in~\cite{AA10}. 

Any $\F_{q}$-linear cyclic codes in $\F_{q}^n$ is an ideal in the residue class 
ring $\F_q[x]/\langle x^n-1\rangle$ (see~\cite[Ch. 4]{HP03} or~\cite[Ch. 7]{MS77}). 
A cyclic code $D$ is a subset of a cyclic code $C$ of equal length over $\F_q$ if and only 
if the generator polynomial of $C$ divides the generator polynomial of $D$. Both polynomials 
divide $x^n-1$.

Since we are also interested in nested pairs of $\F_{r}$-linear
codes over $\F_{q}$, a generalization to the construction of $\F_{r}$-linear
nested cyclic codes over $\F_{q}$ is provided here. Our construction is
a further generalization of~\cite[Th. 14]{CRSS98}.

\begin{definition}
An $(n,r^{l})_{q}$-code $C$ is said to be cyclic $\F_{r}$-linear over
$\F_{q}$ if $C$ is a subspace of the $\F_{r}$-vector space $\F_{q}^{n}$
which is closed under one cyclic shift, \textit{i.e.}, if
$(c_{0},c_{1},\ldots,c_{n-1}) \in C$, then so is $(c_{n-1},c_{0},\ldots,c_{n-2})$.
\end{definition}

Let $\F_{q}$ be the field extension  of $\F_{r}$ of degree
$m$ such that $\F_{q}=\F_{r}(\omega)$. Every polynomial in
$\F_q[x]$ can be uniquely written as
\begin{equation*}
f_0(x)+\omega f_1(x)+\dots+\omega^{m-1} f_{m-1}(x)\text{,}
\end{equation*}
where $f_i(x) \in \F_{r}[x]$ for all $i$.

Given a cyclic $\F_{r}$-linear code $C$ of length $n$ over
$\F_{q}$, we can view the codewords of $C$ as polynomials in $\F_{q}[x]$.
It is often convenient to refer to $C$ as the set
\begin{equation*}
\{ v(x)=v_0+v_1 x+ \ldots +v_{n-1} x^{n-1} : (v_0,v_1,\ldots,v_{n-1}) \in C \}\text{.}
\end{equation*}

Note that, for all $\F_{r} \subseteq \F_{q}$, both $C$ and $\F_{q}[x]/\langle x^n-1 \rangle$ 
are $\F_{r}[x]$-modules under the usual polynomial multiplication together with the rule $x^{n}=1$.

\begin{theorem}\label{thm:cyclic_main}
Let $\F_{q}=\F_{r}(\omega)$ be an extension of degree $m$ over $\F_{r}$.
Any $(n, r^{l})_{q}$-cyclic $\F_{r}$-linear code $C$ over $\F_{q}$ has $m$
generators, and can be represented as an $\F_{r}[x]$-module
\begin{align*}
C = \langle & a_{0,0}(x) + \omega a_{0,1}(x) + \ldots + \omega^{m-1} a_{0,m-1}(x),\\
      & a_{1,0}(x) + \omega a_{1,1}(x) + \ldots + \omega^{m-2} a_{1,m-2}(x), \\
      & \vdots ~~~~~~~~~~~      \vdots ~~~~~~~~~~~~~\iddots \\
      & a_{m-2,0}(x) + \omega a_{m-2,1}(x),\\
      & a_{m-1,0}(x)\rangle,
\end{align*}
where $a_{i,j}(x)\in \F_{r}[x]$ for all $0\leq i \leq m-1$  and $0\leq j \leq i-1$.
Moreover, these polynomials can be chosen such that the following properties hold:
\begin{enumerate}[$i)$]
\item $a_{i,m-1-i}(x) | (x^n-1)$ in $\F_{r}[x]$ for all $0\leq i \leq m-1$.
\item $a_{i ,m-1-i}(x) | \left( a_{i-1,m-1-i}(x)(x^n-1)/a_{i-1,m-i}(x) \right)$ in
$\F_{r}[x]$ for all $1\leq i \leq m-1$.
\item $l=mn-\sum\limits_{i=0}^{m-1}\deg(a_{i,m-1-i}(x))$.
\item The sets
\begin{align*}
& \{a_{i,m-1-i}(x): 0\leq i\leq m-1 \} \text{ and}\\
& \{a_{i,j}(x)\, {\rm mod}\, (a_{m-j,j}(x)) : 0\leq i\leq m-1-j \}
\end{align*}
are unique for all $1\leq j\leq m-1$.
\end{enumerate}
\end{theorem}

\begin{IEEEproof}
See Appendix C.
\end{IEEEproof}
To construct an AQC, from a given cyclic $\F_{r}$-linear
code $C$ over $\F_{q}$ with representation
\begin{align*}
C=\langle & g_0(x) = a_{0,0}(x)+\omega a_{0,1}(x)+\ldots+\omega^{m-1}a_{0,m-1}(x), \\
          & g_1(x) = a_{1,0}(x)+\omega a_{1,1}(x)+\ldots+\omega^{m-2}a_{1,m-2}(x),  \\
          & ~~ \vdots ~~~~~~~~~~~      \vdots ~~~~~~~~~~~~~\iddots \\
          & g_{m-2}(x) = a_{m-2,0}(x)+ \omega a_{m-2,1}(x),\\
          & g_{m-1}(x)=a_{m-1,0}(x)\rangle,
\end{align*}
define $D$ to be the code generated by
\begin{equation*}
\left\lbrace g_0(x)b_0(x),g_1(x)b_1(x),\dots, g_{m-1}(x)b_{m-1}(x)\right\rbrace \text{,}
\end{equation*}
where $b_i(x)$ is a divisor of $(x^n-1)/a_{i,m-1-i}(x)$ for all $0 \leq i \leq m-1$.
$D$ is a cyclic $\F_{r}$-linear subcode of $C$.

\subsection{Construction from Group Character Codes}\label{subsec:GC}
Group character (GC) codes were introduced in~\cite{DKL00} based on elementary
abelian $2$-groups and were further generalized in~\cite{Ling04} to include
the case where the group is $\left(\Z/t\Z \right)^{l}$ for $l, t \in \N$. 
We use the definitions and results in~\cite{Ling04} for generality.

The elements of $\left(\Z/t\Z \right)^{l}$ can be written as 
$(a_{1},\ldots,a_{l})$ where $0 \leq a_{i} \leq t-1$ for $1 \leq i \leq l$.
Let $||a||=\sum_{i=1}^{l} a_{i} \in \Z$. Note that $0 \leq ||a|| \leq (t-1)l$ for
all $a \in \left(\Z/t\Z \right)^{l}$.

Let $r$ be an integer such that $0 \leq r < l(t-1)$ and let the set $\calX (r,l;t)$ be given by
\begin{equation}\label{eq:defX}
 \calX(r,l;t) = \{ a \in \left(\Z/t\Z \right)^{l}: ||a|| > r\}\text{.}
\end{equation}

\begin{definition}\label{defGCCodes}
Let $\F_{q}$ be a finite field with $t | (q-1)$ and let $f_{0},f_{1},\ldots f_{t^{l}-1}$ be 
the group characters from $\left(\Z/t\Z \right)^{l}$ to $\F_{q}\setminus\{0\}$. Let $\c = \left( c_{0},c_{1},\ldots,c_{t^{l}-1} \right)$ be a vector in $\F_{q}^{t^{l}}$. 
Let $C_{q}(r,l;t)$ denote the $q$-ary code
\begin{equation}\label{eq:GCCodes}
 C_{q}(r,l;t)=\left\lbrace \c : \sum_{j=0}^{t^{l}-1} c_{j}f_{j}(x)=0
\text{ for all } x \in \calX(r,l;t) \right\rbrace \text{.}
\end{equation}
\end{definition}

The properties of $C_{q}(r,l;t)$ are known.
\begin{theorem}\cite[Th. 8]{Ling04}\label{thm:GCProperties}
Writing $r=a(t-1)+b$, where $0 \leq b \leq t-2$, the code $C_{q}(r,l;t)$ has
parameters
\begin{equation}\label{eq:properties}
[t^{l},k_{l}(r),(t-b)t^{l-1-a}]_{q}\text{,}
\end{equation}
where
\begin{equation*}
 k_{l}(r)=\sum_{i=0}^{r} \sum_{j=0}^{l} (-1)^{j}\binom{l}{j}\binom{l-1+i-jt}{t-1} \text{.}
\end{equation*}
\end{theorem}

The nestedness condition can be deduced directly from (\ref{eq:defX}) and (\ref{eq:GCCodes}).
\begin{lemma}\label{lem:nested}
If $0 \leq r_{1} \leq r_{2} < l(t-1)$, then $C_{q}(r_{1},l;t)
\subseteq C_{q}(r_{2},l;t)$.
\end{lemma}

\begin{theorem}\cite[Th. 10]{Ling04}\label{thm:dualGC}
The Euclidean dual $(C_{q}(r,l;t))^{\perp_{\E}}$ of
$C_{q}(r,l;t)$ is monomially equivalent\footnote{A discussion on code equivalence can 
be found in~\cite[Sects. 1.6 and 1.7]{HP03}.} to $C_{q}(l(t-1)-1-r,l;t)$.
\end{theorem}
Hence, $d\left((C_{q}(r,l;t))^{\perp_{\E}} \right)$ can be computed explicitly.

\begin{theorem}\label{thm:GCquantum}
Let $0 \leq r_{1} \leq r_{2} < l(t-1)$. Let $a, b, \gamma, \delta, k,
d_{1}$ and $d_{2}$ be nonnegative integers satisfying
\begin{align*}
 r_{2} &= a (t-1) + b \text{ where } 0 \leq b < t-1 \text{,}\\
 l(t-1)-1-r_{1} &= \gamma (t-1) + \delta \text{ with } 0 \leq \delta < t-1 \text{,} \\
 k &= k_{l} (r_{2}) - k_{l} (r_{1}) \text{,}\\
 d_{1} & = (t - \delta)t^{l-1-\gamma}\text{, and} \\
 d_{2} & = (t-b)t^{l-1-a} \text{.}
\end{align*}
Then there exists an asymmetric stabilizer code $Q$ with parameters
$[[t^{l},k,d_{2}/d_{1}]]_{q}$.
\end{theorem}
\begin{IEEEproof}
Use the nested pair $C_{q}(r_{1},l;t) \subseteq C_{q}(r_{2},l;t)$ in Theorem~\ref{thm:standardCSS}. 
Combining Theorem~\ref{thm:dualGC} and (\ref{eq:properties}), we get
\begin{equation*}
d\left((C_{q}(r_{1},l;t))^{\perp_{\E}} \right)
= (t - \delta)t^{l-1-\gamma} =d_{1}  \text{,}
\end{equation*}
if we write $l(t-1)-1-r_{1} =
\gamma (t-1) + \delta$ with $0 \leq \delta < t-1 $.
The other values are clear.
\end{IEEEproof}

\subsection{BKLC construction}\label{subsec:BKLC}
Let us start with the following result.
\begin{theorem}\label{thm:spectrumn}
Let $C$ be a linear $[n,k,d]_{q}$-code. If $C$ has a 
codeword $\v$ such that $\wt(\v)=n$, then there exists 
an $[[n,k-1,d/2]]_{q}$-code $Q$.
\end{theorem}

\begin{IEEEproof}
Construct a code $D:=\{\lambda \v : \lambda \in \F_{q}\} \subset C$. 
Since $D$ is MDS, $D^{\perp}$ is also MDS with parameters $[n,n-1,2]_{q}$. 
Setting $C_{1}=D^{\perp}$ and $C_{2}=C$ in Theorem~\ref{thm:standardCSS} completes the proof.
\end{IEEEproof}

An obvious strategy is to identify the best-known linear codes stored in the
database of MAGMA that contain a codeword of weight equal to the length
$n$ for small fields $q \in \{2,3,4,5,7,8,9\}$. We call this construction the
\textit{BKLC construction}.

Note that sometimes the database does not contain a linear code of specified
length $n$ and dimension $k$ satisfying the required condition since this
specific requirement has not been recognized as important before. 
This in no way excludes the possibility of the existence of a linear code that 
has a codeword of weight $n$.

\section{Linear Programming Bounds}\label{sec:LP}

This section details the set-up and the implementation of the linear programming (LP) bounds (more precisely, systems of linear inequalities) 
that we use to derive the upper bound for $k = \log_q(K)$ (see~\cite{SKR09} for an earlier attempt in the binary case). 
Again, let $*$ stand for any one of the Euclidean, trace Euclidean, Hermitian, and trace Hermitian inner products. 
In fact, without loss of generality, $*$ can be taken as the trace Euclidean inner product based on Fig. 1.

From Section~\ref{sec:part_one}, given $q=r^{m},n,k,d_{x},d_{z}$, a pure CSS-like $[[n,k,d_{z}/d_{x}]]_q$ code exists if and 
only if there exists  a pair $C_{1},C_{2}$ of $\F_{r}$-linear codes over $\F_{q}$ such that $C_{1}^{\perp_*}\subset C_{2}$,  
$k=\log_{q} \left(\frac{|C_{2}|}{|C_{1}^{\perp_{*}}|} \right)$ with $d_{x}=d_{1}$ and $d_{z}=d_{2}$.

If LP rules out the existence of such a pair, a negative certificate is issued. Otherwise, the process indicates 
the values of $k$ which cannot be ruled out and the parameters of the (hypothetical) pair $C_{1}$ and $C_{2}$ 
giving such $k$. This information is useful when we try to come up with some {\it ad hoc} constructions yielding good codes 
as illustrated, {\it e.g.}, in Subsection~\ref{subsec:AdHoc}

For $0 \leq j \leq n$, let $A_{j}$ and $B_{j}$ be, respectively, the number of codewords of weight $j$ in $C_{2}$ and $C_{1}$. 
The corresponding numbers $A_{j}^{\perp_{*}}$ and $B_{j}^{\perp_{*}}$ of their respective duals are given by (\ref{eq:Krawteq}). 
One can write column vectors $B,A, B^{\perp_{*}}$, and $A^{\perp_{*}}$, each having $n+1$ entries to represent the 
weight distributions of $C_{1}, C_{2}$, and of their duals, respectively. Introduce the matrix $K$ in the space of real 
$(n+1) \times (n+1)$ matrices $\cM_{n+1} (\R)$ with $K_{j,i} = K^{n,q}_{j}(i) \in \Z$ from (\ref{eq:Krawt}).

Since $C_{1}^{\perp_{*}} \subset C_{2}$, we obtain $C_{2}^{\perp_{*}} \subset C_{1}$ by taking their duals. Given these two 
pairs of nested codes one can follow Delsarte's approach~\cite{Del72} to derive bounds for $|C_{i}|$ and $|C_{i}^{\perp_{*}}|$ 
for $i=1,2$. For an arbitrary code $C \subseteq \F_{q}^{n}$ of minimum distance $d$ having $\calW$ as a feasible column 
vector representing its weight distribution,
\begin{equation}\label{eq:DelBound}
|C| \leq \max_{\calW \geq \0} \sum_{i=0}^{n} \calW_{i} \text{ such that } \mathcal{D}:= K \calW \geq \0
\end{equation}
provided that $\calW_{0}=1$ and $\calW_{s}=0$ for $1 \leq s \leq d-1$.

If $C$ is $\F_{r}$-linear, let $\ddual$ be the minimum distance of $C^{\perp_{*}}$ and $\Wdual$ be a feasible vector 
representation of its weight distribution. One can then write, for $|C|=q^{l}$, 
$$ K \calW = q^{l} \Wdual \geq \0$$ 
given that $\calW_{0}^{\perp_{*}}=1$ and $\calW_{t}^{\perp_{*}}=0$ for $1 \leq t \leq \ddual-1$. This means that (\ref{eq:DelBound}) can be 
improved by adding the constraints that $\mathcal{D}_{i}=0$ for $1 \leq i \leq \ddual-1$.

From here on we assume that $C$ is $\F_{r}$-linear. Let $D(d):=\lfloor \log_{r} \left(\max |C| \text{ in (\ref{eq:DelBound})}\right) \rfloor$ 
be the largest possible $\F_{r}$-dimension of $C$ under Delsarte's bound. Then any such code $C$ of minimum distance $d$ 
satisfies $|C| \leq r^{D(d)}$. In a similar fashion, let $D(d,\ddual)$ denote $\lfloor \log_{r} \left(\max |C|\right)\rfloor$ when $C$ has 
minimum distance $d$ and $C^{\perp_{*}}$ has minimum distance $\ddual$. Under this improved bound, as $|C| =r^{ml} \leq r^{D(d,\ddual)}$, 
one has $|C^{\perp_{*}}| \geq r^{mn - D(d,\ddual)}$. Thus,\footnote{The above and what follows do not depend on the particular 
nature of the bound $D(d,\ddual)$.}
\begin{equation*}
	D(\ddual,d) \geq \dimr (C^{\perp_{*}}) \geq mn - D(d,\ddual) \text{.}
\end{equation*}
Dually, we get
\begin{equation*}\label{eq:dim1}
	D(d,\ddual) \geq \dimr (C) \geq mn - D(\ddual,d) \text{.}
\end{equation*}

To limit our search space, we need to establish feasible values for the pair $(k,k')$ to be used as part of the 
input to establish the LP bound. Let $|C_{1}|=q^{k+k'}$ and $|C_{2}|=q^{n-k'}$ for $mk,mk' \in \Z$. Let $d_{x}=d(C_{1})$ 
and $d_{z}=d(C_{2})$ be given. Let $\alpha :=D(d_{x},d_{z})$ and $\beta:=D(d_{z},d_{x})$. Since $C_{1}^{\perp_{*}} \subset C_{2}$, 
the pair of codes $\left(C_{1},C_{1}^{\perp_{*}}\right)$ satisfies $d(C_{1})=d_{x}$ and $d(C_{1}^{\perp_{*}}) \geq d_{z}$. This gives
 \begin{align*}
 	& \alpha \geq \dimr (C_{1}) = m(k+k') \geq mn-\beta \text{ and}\\
 	& \beta  \geq \dimr (C_{1^{\perp_{*}}}) = m(n-k-k') \geq mn-\alpha \text{.}
 \end{align*}
 
Looking at the duals, since $C_{2}^{\perp_{*}} \subset C_{1}$, one has $d(C_{2}^{\perp_{*}}) \geq d_{x}$. The pair of codes 
$\left(C_{2},C_{2}^{\perp_{*}}\right)$ satisfies $d(C_{2})=d_{z}$ and $d(C_{2}^{\perp_{*}}) \geq d_{x}$. Hence,
\begin{align*}
	&\beta \geq \dimr (C_{2}) = m(n-k') \geq mn-\alpha \text{ and}\\
	&\alpha \geq \dimr (C_{2}^{\perp_{*}}) = m k' \geq mn-\beta \text{.}
\end{align*}

These sets of inequalities are equivalent to the system
\begin{equation}\label{eq:system}
\begin{cases}
	m(k+k') \leq \alpha \\
	mk' \geq mn - \beta
\end{cases}\text{.}
\end{equation}

We add $mk \geq 1$ since $C_{2}^{\perp_{*}}$ is a strict subset of $C_{1}$ and $0 < k' < n-k$ to ensure 
$d_z,d_x > 1$ since our AQC $Q$ should have both $X$-error and $Z$-error detection capability. 

Drawing the picture for feasible $(mk,mk')$, the possible pairs must correspond to the integer points in the gray triangle.

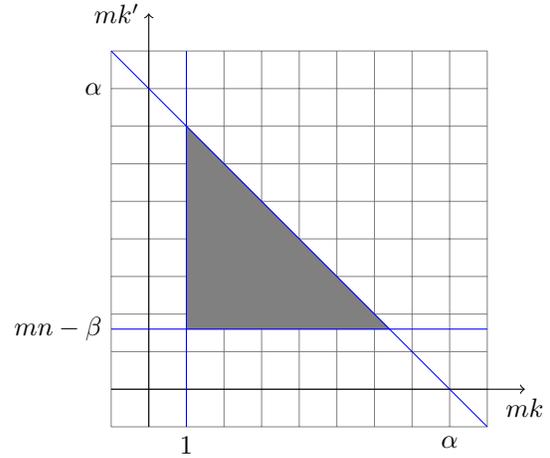
\begin{tikzpicture}[scale=1]\label{pic}
\filldraw[color=gray] (0.5,0.8) -- (0.5,3.5) -- (3.2,0.8)  -- cycle;
\draw[help lines, step=0.5] (-0.5,-0.5) grid (4.5,4.5);
\draw[->,color=black] (-0.5,0) -- (5,0) node[below] {$mk$};
\draw[->,color=black] (0,-0.5) -- (0,5) node[left] {$mk'$};

\draw [ color=blue ] (0.5,-0.5) -- (0.5,4.5);
\draw[blue] plot[smooth, domain=-0.5:4.5] (\x,{-1*\x+4});
\draw[blue] plot[smooth, domain=-0.5:4.5] (\x,0.8); 

\draw [ color=black ] (-0.5,0.8) node[left]{$mn-\beta$}   (-0.5,4) node[left]{$\alpha$};
\draw [ color=black ] (0.5,0-0.5) node[below]{$1$}   (4,-0.5) node[below]{$\alpha$};
\end{tikzpicture}

While many tuples  $(n,q,d_{x},d_{z})$ are ruled out this way, there are situations when there will 
be feasible $(mk,mk')$. 

\begin{example}
Consider $(n,q,d_x,d_z)=(7,4,5,2)$ for $m=2$. By Delsarte's bound, it is known (see~\cite{BBKO01} for more details) 
that the largest sizes of $\F_{4}$-codes with $d=5$ and $d=2$ are bounded above by, respectively, $40$ and $4096$. 
In this case, $\alpha=5=\lfloor\log_2(40)\rfloor$ and $\beta=12=\log_2(4096)$. Thus, the gray triangle containing all 
six possible $(mk,mk')$ values has vertices $(1,2)$, $(1,4)$, and $(3,2)$.
\end{example}

Once feasible $(mk,mk')$ values are found, one can prepare the input tuple $(n,q,k,k',d_{x},d_{z})$ 
for the {\it formal} LP whose objective function\footnote{In fact, any linear function can be chosen.} is to maximize
\begin{equation*}
\sum_{j=1}^{d_{z}-1} A_{j} \text{,}
\end{equation*}
subject to the following constraints:
\begin{enumerate}
 \item $A_{0}=B_{0}=A_{0}^{\perp_{*}}=B_{0}^{\perp_{*}}= 1$, \label{ieq:first}
 \item $A_{j} = 0$ for $1 \leq j < d(C_{2})$ and $A_{j} \geq 0$ for $j \geq d(C_{2})$,
 \item $B_{j} = 0$ for $1 \leq j < d(C_{1})$ and $B_{j} \geq 0$ for $j \geq d(C_{1})$,
 \item $A_{j}^{\perp_{*}} = 0$ for $1 \leq j < d(C_{2}^{\perp_{*}})$ and $A_{j}^{\perp_{*}} \geq 0$ for $j \geq d(C_{2}^{\perp_{*}})$,
 \item $B_{j}^{\perp_{*}} = 0$ for $1 \leq j < d(C_{1}^{\perp_{*}})$ and $B_{j}^{\perp_{*}} \geq 0$ for $j \geq d(C_{1}^{\perp_{*}})$,
 \item $K A^{\perp_{*}} = q^{k'} A$,
 \item $K B^{\perp_{*}} = q^{n-k-k'} B$,
 \item $A_{j}=B_{j}^{\perp_{*}}$ for $0 \leq j \leq d_{z}-1$, $A_{d_z} > B_{d_z}^{\perp_{*}}$, 
and $A_{j} \geq B_{j}^{\perp_{*}}$ for all $d_{z} < j \leq n$,
 \item $B_{j}=A_{j}^{\perp_{*}}$ for $0 \leq j \leq d_{x}-1$, $B_{d_x} > A_{d_x}^{\perp_{*}}$, 
and $B_{j} \geq A_{j}^{\perp_{*}}$ for all $d_{x} < j \leq n$.\label{ieq:last}
\end{enumerate}

Constraints 6 and 7 come from combining (\ref{eq:Krawteq}) and the fact that $K^{2}=q^{n}I$ where $I$ is the identity matrix. 
The last two constraints take care of the purity assumption that $\wt(C_{2} \setminus C_{1}^{\perp_{*}}) = d(C_{2})$ and 
$\wt(C_{1} \setminus C_{2}^{\perp_{*}}) = d(C_{1})$.

The latter LP rules out, for instance, the tuple $(n,q,k,k',d_x,d_z)=(6,2,2,1,3,2)$, which is not ruled out by the gray triangle above. 
Indeed, for $(n,q,d_x,d_z)=(6,2,3,2)$, the integer points $(k,k')$ in the triangle are $(1,1)$, $(1,2)$, and $(2,1)$. For the first 
two tuples, the LP is feasible. For the third, one can compute a {\it Farkas-like} 
certificate of infeasibility for the system \ref{ieq:first})--\ref{ieq:last}) as follows. 

After reordering the constraints and multiplying some of them, if necessary, by $-1$, the system can be
rewritten as
\begin{equation}\label{eq:Monetwo}
M_1\cdot \binom{A}{B}=\mathbf{r} \geq \mathbf{0}, \ M_2\cdot \binom{A}{B}\geq \mathbf{0},
\end{equation}
where $M_1$, $M_2$ are matrices with $2n+2$ columns each and $\mathbf{0}\neq \mathbf{r}$ is a nonnegative vector. 

One then tries to find a vector $\mathbf{s}=(\mathbf{s}_1 \mathbf{s}_2)$ satisfying 
$(M_1^\intercal M_2^\intercal)\binom{\mathbf{s}_1}{\mathbf{s}_2}\leq \mathbf{0}$ such that $\mathbf{s}_2\geq\mathbf{0}$ and $\mathbf{s}_1^\intercal \mathbf{r}>0$.
It follows from an appropriate form of the Farkas Lemma that such an $\mathbf{s}$ exists if and only if the system \ref{ieq:first})--\ref{ieq:last})
is infeasible. To see sufficiency, note that $\mathbf{s}^\intercal \binom{M_1}{M_2}\binom{A}{B}\leq 0$, 
whereas $\mathbf{s}^\intercal \binom{\mathbf{r}}{\mathbf{0}}>0$, a contradiction.

The vector $\mathbf{s}$ certifying infeasibility can be found by linear programming. The details of such a computation for 
$(n,q,k,k',d_x,d_z)=(6,2,2,1,3,2)$ is in Appendix D.

\section{Good Pure CSS-like AQCs based on Linear Programming Bound}\label{sec:Optimal}
Based on the LP bound, this section presents good AQCs derived from the 
nested pairs of classical codes constructed by the methods outlined in Section~\ref{sec:part_three}.

By Proposition~\ref{prop:Sorth}, good AQCs with $K=1$ and $d_{z}=d_{x}$ can 
be derived from self-dual codes having the largest possible minimum distance.
Lists of extremal and optimal self-dual codes over various finite fields can
be found in~\cite[Ch. 11]{NRS06}. More recent results are available in~\cite{GG09},~\cite{GKL08} as 
well as prominent references therein. The parameters of the AQCs that can be 
derived from these extremal or optimal self-dual codes via CSS-like constructions 
can be computed easily. In the case of $q=4$, for example,~\cite[Table I]{ELS10} 
provides the most updated list. Henceforth, we consider AQCs with $K > 1$.

Among the best pure AQCs is of course the class of codes reaching the equality 
of the quantum Singleton bound $K \leq q^{n-d_{z}-d_{x}+2}$. Such codes are 
referred to as AQMDS codes whose full treatment can be found in~\cite{EJKL11}. 
Assuming the validity of the classical MDS conjecture, the lengths of pure AQMDS codes 
are bounded above roughly by $q$. It is of interest, therefore, to identify the best 
possible pure CSS-like AQCs for lengths beyond the possible values for the MDS type.

According to the LP bound, Table~\ref{table:good} gives a criterion for the goodness of the constructed AQCs.

\begin{table}
\caption{Measure of Goodness}
\label{table:good}
\centering
\begin{tabular}{| c | p{6.2cm} |}
\hline
\textbf{Label} & \textbf{Description} \\
\hline
\textit{Optimal} &	The LP bound for $k$ is reached.\vspace{0.1cm}\\

\textit{BeOpLin} &	The pair of nested subfield linear codes yields better $k$ than the LP bound value when $\F_{q}$-linearity is imposed.\vspace{0.1cm} \\
  
\textit{OpLin}   &	The LP bound with $\F_{q}$-linearity required is attained.\vspace{0.1cm}\\

\textit{ROpLin}  &	The pair of nested subfield linear codes yields the LP bound value when $\F_{q}$-linearity is imposed.\\
\hline
\end{tabular}
\end{table}

To present our findings in as concise a manner as possible, we separate the 
tables of good pure AQCs according to the fields. When $q$ is a prime, only $\F_{q}$-linear 
pairs are possible. The results are presented in Subsection~\ref{subsec:TabPrimes}. 

When $\F_q$ is a nontrivial extension of $\F_{p}$, then we need to consider also the case 
where the pairs consist of subfield linear codes. For $q \in \{4,8,9\}$, we differentiate 
between the strictly $\F_{q}$-linear cases and the $\F_{r}$-linear cases. 
AQCs from $\F_{r}$-linear construction beating the best that the strictly 
$\F_{q}$-linear construction can achieve are listed as well to highlight the gain that 
we get from going non-$\F_{q}$-linear. Subsection~\ref{subsec:TabPowerPrimes} presents the tables.

In both subsections, the tables are ordered according to $n, d_x$ and $d_z$. The following shorthands are 
used to distinguish the types of construction:
\begin{enumerate}
 \item ACC stands for $\F_{r}$-linear but not $\F_{q}$-linear nested cyclic pair of codes where 
at least one of the codes in the pair is not $\F_{q}$-linear.
 \item AH stands for an \textit{ad hoc} pair of codes. Their explicit construction will be 
provided in detail in Subsection~\ref{subsec:AdHoc}.
 \item BC stands for a nested pair of codes where the supercode is taken from the
MAGMA's database of best-known linear codes having a codeword $\v$ with $\wt(\v)=n$.
 \item CC stands for $\F_{q}$-linear nested cyclic pair of codes.
 \item GC stands for $\F_{q}$-linear nested pair of group character codes.
 \item For $q=4$, the type SO refers to an AQC which is derived from a self-orthogonal 
code $C$ discussed in~\cite[Table II, Sect. VII]{ELS10}.
\end{enumerate}

\subsection{Tables of Optimal Pure Asymmetric CSS Codes for $q \in \{2,3,5,7\}$}\label{subsec:TabPrimes}
The lists of optimal pure AQCs for $q \in \{2,3,5,7 \}$ are given, respectively for each $q$, 
in Tables~\ref{table:F2Optimal},~\ref{table:F3Optimal},~\ref{table:F5Optimal}, and~\ref{table:F7Optimal}.

For each $q$, the table is then followed by the table giving the explicit pairs of cyclic 
codes $D \subset C$ yielding them. To save space, the generator polynomials of the codes $C$ and $D$ are presented in an 
abbreviated form. The generator polynomial $g(x)$ of the $[n,k_{C}]_{q}$-code $C$ is written as $g=(c_{0} c_{1} \ldots c_{n-k_{C}})$ 
instead of as the polynomial $g(x)=c_{0}+c_{1}x+\ldots+c_{n-k_{C}}x^{n-k_{C}}$. Since $g(x)$ divides the 
generator polynomial of $D$, we write the latter as $(d_{0} d_{1} \ldots d_{k_{C}-k_{D}})g$ where $k_{D}$ 
is the dimension of $D$. The details on the explicit cyclic pairs can be found in 
Tables~\ref{table:F2Cyclic},~\ref{table:F3Cyclic},~\ref{table:F5Cyclic}, and~\ref{table:F7Cyclic}.

\begin{table*}
\caption{Optimal Pure Asymmetric CSS Codes over $\F_{2}$}
\label{table:F2Optimal}
\centering
\begin{tabular}{| c  l  c || c  l  c || c  l  c || c  l  c |}
\hline
\textbf{No.} & \textbf{AQC $Q$} & \textbf{Type}
& \textbf{No.} & \textbf{AQC $Q$} & \textbf{Type}
& \textbf{No.} & \textbf{AQC $Q$} & \textbf{Type}
& \textbf{No.} & \textbf{AQC $Q$} & \textbf{Type} \\
\hline
1 & $[[4,1,2/2]]_{2}$ & AH &  22  & $[[15,4,4/4]]_{2}$ & CC &  43  & $[[24,18,3/2]]_{2}$ & BC & 64 & $[[31,15,6/3]]_{2}$ & CC \\
2 & $[[5,2,2/2]]_{2}$ & AH &  23  & $[[15,2,5/4]]_{2}$ & CC & 44  & $[[24,17,4/2]]_{2}$ & BC & 65 & $[[31,6,11/3]]_{2}$ & CC \\
3 & $[[6,1,3/2]]_{2}$ & CC &  24  & $[[16,10,4/2]]_{2}$& BC & 45  & $[[24,13,6/2]]_{2}$ & BC & 66 & $[[31,5,12/3]]_{2}$ & CC \\
4 & $[[7,3,3/2]]_{2}$ & CC &  25  & $[[16,4,8/2]]_{2}$ & BC & 46  & $[[24,11,8/2]]_{2}$ & BC & 67 & $[[31,1,15/3]]_{2}$ & CC \\
5 & $[[7,1,3/3]]_{2}$ & CC &  26  & $[[17,8,5/2]]_{2}$ & CC & 47  & $[[26,1,13/2]]_{2}$ & CC & 68 & $[[31,15,5/4]]_{2}$ & CC \\

6 & $[[8,3,4/2]]_{2}$ & BC &  27  & $[[17,1,5/5]]_{2}$ & CC & 48  & $[[27,13,7/2]]_{2}$ & BC & 69 & $[[31,5,11/4]]_{2}$ & CC \\
7 & $[[10,1,5/2]]_{2}$ & CC & 28  & $[[18,8,6/2]]_{2}$ & BC & 49  & $[[28,22,3/2]]_{2}$ & BC & 70 & $[[31,11,5/5]]_{2}$ & CC \\
\hline
\end{tabular}
\end{table*}

\begin{table*}
\caption{Nested Pairs of Cyclic Codes over $\F_{2}$ Yielding Optimal Asymmetric CSS Codes in Table~\ref{table:F2Optimal}}
\label{table:F2Cyclic}
\centering
\begin{tabular}{| l l l || l l l |}
\hline
\textbf{$C$ and $D$}
& \textbf{Generator Polynomials} & \textbf{AQC $Q$} 
& \textbf{$C$ and $D$}
& \textbf{Generator Polynomials} & \textbf{AQC $Q$}\\
\hline
$[6,5,2]_2$ & $g=(1 1)$ & $[[6,1,3/2]]_{2}$ &   $[21,12,5]_2$& $g=(1 1 1 0 1 1 0 0 1 1)$ & $[[21,3,5/5]]_{2}$ \\
$[6,4,2]_2$ & $(1 1)g$  &                   &   $[21,9,8]_2$ &  $(1 0 0 1)g$             &  \\

$[7,4,3]_2$ & $g=(1 0 1 1)$&$[[7,3,3/2]]_{2}$&   $[21,12,5]_2$& $g=(1 1 1 0 1 1 0 0 1 1)$ & $[[21,2,6/5]]_{2}$\\
$[7,1,7]_2$ & $(1 1 0 1)g$  &                   & $[21,10,5]_2$& $(1 1 1)g$ & \\

$[7,4,3]_2$ & $g=(1 0 1 1)$ & $[[7,1,3/3]]_{2}$  &  $[22,21,2]_2$& $g=(1 1)$ & $[[22,1,11/2]]_{2}$ \\ 
$[7,3,4]_2$ & $(1 1)g$ &                         & $[22,20,2]_2$& $(1 1)g$ & \\

$[10,9,2]_2$ &$g=(1 1)$&$[[10,1,5/2]]_{2}$ &  $[23,12,7]_2$& $g=(1 1 0 0 0 1 1 1 0 1 0 1)$ & $[[23,11,7/2]]_{2}$\\
\hline
\end{tabular}
\end{table*}
\begin{table*}
\caption{Optimal Pure Asymmetric CSS Codes over $\F_{3}$}
\label{table:F3Optimal}
\centering
\begin{tabular}{| c  l  c || c  l  c || c  l  c || c  l  c |}
\hline
\textbf{No.} & \textbf{AQC $Q$} & \textbf{Type}
& \textbf{No.} & \textbf{AQC $Q$} & \textbf{Type}
& \textbf{No.} & \textbf{AQC $Q$} & \textbf{Type}
& \textbf{No.} & \textbf{AQC $Q$} & \textbf{Type} \\
\hline
1 & $[[6,2,3/2]]_{3}$ & CC & 49 & $[[17,11,4/2]]_{3}$ & BC & 97 & $[[24,16,5/2]]_{3}$  & BC & 145 & $[[30,25,3/2]]_{3}$ & BC \\
2 & $[[6,1,4/2]]_{3}$ & CC & 50 & $[[17,9,5/2]]_{3}$  & BC & 98 & $[[24,15,6/2]]_{3}$  & BC & 146 & $[[30,23,4/2]]_{3}$ & BC \\
3 & $[[7,3,3/2]]_{3}$ & BC & 51 & $[[17,8,6/2]]_{3}$ & BC & 99 & $[[24,11,9/2]]_{3}$  & BC & 147 & $[[30,4,18/2]]_{3}$ & BC \\
4 & $[[7,2,4/2]]_{3}$ & BC & 52 & $[[17,6,7/2]]_{3}$ & BC & 100 & $[[24,6,12/2]]_{3}$ & BC & 148 & $[[30,1,22/2]]_{3}$ & BC \\
5 & $[[8,4,3/2]]_{3}$ & CC & 53 & $[[17,5,8/2]]_{3}$ & BC & 101 & $[[24,3,15/2]]_{3}$ & CC & 149 & $[[31,26,3/2]]_{3}$ & BC \\

6  & $[[8,3,4/2]]_{3}$ & CC,GC & 54 & $[[18,13,3/2]]_{3}$ & BC & 102 & $[[24,16,3/3]]_{3}$ & CC & 150 & $[[31,24,4/2]]_{3}$ & BC \\
7  & $[[8,2,5/2]]_{3}$ & CC &     55 & $[[18,12,4/2]]_{3}$ & BC & 103 & $[[24,14,4/3]]_{3}$ & CC & 151 & $[[31,1,23/2]]_{3}$ & BC \\
\hline
\end{tabular}
\end{table*}
\begin{table*}
\caption{Nested Pairs of Cyclic Codes over $\F_{3}$ Yielding Optimal Asymmetric CSS Codes in Table~\ref{table:F3Optimal}}
\label{table:F3Cyclic}
\centering
\begin{tabular}{| l l l || l l l |}
\hline
\textbf{$C$ and $D$} & \textbf{Generator Polynomials} & \textbf{AQC $Q$} 
& \textbf{$C$ and $D$}& \textbf{Generator Polynomials} & \textbf{AQC $Q$}\\
\hline
$[6,5,2]_3$&   $ g=(2 1)$ & $[[6,2,3/2]]_{3}$ & $[22,16,4]_3$& $ g=(2 1 0 0 2 1 1)$& $[[22,10,4/4]]_{3}$\\
$[6,3,3]_3$&   $(1 2 1) g$ &  & $[22,6,12]_3$& $(2 1 2 2 1 2 2 2 1 1 1) g$&\\

$[6,5,2]_3$&   $ g=(2 1)$& $[[6,1,4/2]]_{3}$ & $[22,12,7]_3$& $ g=(2 2 2 1 1 1 2 1 1 2 1)$& $[[22,6,7/4]]_{3}$ \\
$[6,4,2]_3$&   $(2 1) g$& & $[22,6,12]_3$& $(2 2 1 0 0 2 1) g$&\\

$[8,7,2]_3$&   $ g=(1 1)$& $[[8,4,3/2]]_{3}$ & $[22,12,7]_3$& $ g=(2 2 2 1 1 1 2 1 1 2 1)$& $[[22,2,7/7]]_{3}$ \\
$[8,3,5]_3$&   $(2 2 0 2 1) g$& & $[22,10,9]_3$& $(2 0 1) g$&\\

$[8,7,2]_3$&   $ g=(1 1)$& $[[8,3,4/2]]_{3}$ & $[23,12,8]_3$& $ g=(2 2 2 1 1 0 2 0 2 0 0 1)$& $[[23,1,8/8]]_{3}$ \\
\hline
\end{tabular}
\end{table*}
\begin{table*}
\caption{Optimal Pure Asymmetric CSS Codes over $\F_{5}$}
\label{table:F5Optimal}
\centering
\begin{tabular}{| c  l  c || c  l  c || c  l  c || c  l  c |}
\hline
\textbf{No.} & \textbf{AQC $Q$} & \textbf{Type}
& \textbf{No.} & \textbf{AQC $Q$} & \textbf{Type}
& \textbf{No.} & \textbf{AQC $Q$} & \textbf{Type}
& \textbf{No.} & \textbf{AQC $Q$} & \textbf{Type} \\
\hline
1 & $[[6,2,4/2]]_{5}$ & CC & 48 & $[[13,5,4/4]]_{5}$ & CC    & 95 & $[[19,2,14/2]]_{5}$ & BC & 142 & $[[26,21,4/2]]_{5}$ & BC \\
2 & $[[7,2,4/2]]_{5}$ & BC & 49 & $[[13,1,7/4]]_{5}$ & CC    & 96 & $[[20,16,3/2]]_{5}$ & CC & 143 & $[[26,19,5/2]]_{5}$ & BC \\
3 & $[[7,3,3/2]]_{5}$ & BC & 50 & $[[14,10,3/2]]_{5}$ & BC   & 97 & $[[20,15,4/2]]_{5}$ & BC & 144 & $[[26,18,6/2]]_{5}$ & BC \\
4 & $[[8,4,3/2]]_{5}$ & CC & 51 & $[[14,9,4/2]]_{5}$ & BC    & 98 & $[[20,13,5/2]]_{5}$ & BC & 145 & $[[26,6,16/2]]_{5}$ & BC \\
5 & $[[8,3,4/2]]_{5}$ & BC,GC & 52 & $[[14,7,5/2]]_{5}$ & BC & 99 & $[[20,12,6/2]]_{5}$ & BC & 146 & $[[27,23,3/2]]_{5}$ & BC \\

6 & $[[8,2,5/2]]_{5}$ & BC & 53 & $[[14,6,6/2]]_{5}$ & BC   & 100 & $[[20,5,12/2]]_{5}$ & BC & 147 & $[[27,21,4/2]]_{5}$ & BC \\
7 & $[[8,1,6/2]]_{5}$ & CC & 54 & $[[14,5,7/2]]_{5}$ & BC   & 101 & $[[20,4,13/2]]_{5}$ & BC & 148 & $[[27,20,5/2]]_{5}$ & BC \\
\hline
\end{tabular}
\end{table*}
\begin{table*}
\caption{Nested Pairs of Cyclic Codes over $\F_{5}$ Yielding Optimal Asymmetric CSS Codes in Table~\ref{table:F5Optimal}}
\label{table:F5Cyclic}
\centering
\begin{tabular}{| l l l || l l l |}
\hline
\textbf{$C$ and $D$} & \textbf{Generator Polynomials} & \textbf{AQC $Q$} 
& \textbf{$C$ and $D$}& \textbf{Generator Polynomials} & \textbf{AQC $Q$}\\
\hline 
$[6,3,4]_5$ &   $g=(1 2 2 1)$ & $[[6,2,4/2]]_{5}$ & $[13,9,4]_5$ &  $g=(1 1 4 1 1)$ & $[[13,1,7/4]]_{5}$ \\
$[6,1,6]_5$ &   $(1 4 1)g$    &                   & $[13,8,4]_5$ &  $(4 1)g$        & \\

$[8,7,2]_5$ &   $g=(1 1)$      & $[[8,4,3/2]]_{5}$ & $[15,2,12]_5$ & $g=(4 3 2 1 0 4 3 2 1 0 4 3 2 1)$ & $[[15,1,12/2]]_{5}$ \\
$[8,3,4]_5$ &   $(1 1 0 2 1)g$ &                   & $[15,1,15]_5$ & $(4 1)g$ & \\ 

$[8,7,2]_5$ &   $g=(1 1)$    & $[[8,3,4/2]]_{5}$ & $[20,17,3]_5$ & $g=(4 0 2 1)$ & $[[20,16,3/2]]_{5}$\\ 
$[8,4,4]_5$ &   $(4 2 2 1)g$ &                   & $[20,1,20]_5$ & $(1 4 3 3 1 3 1 1 1 2 4 4 1 1 2 2 1)g$ & \\  

$[8,7,2]_5$ &   $g=(1 1)$ & $[[8,1,6/2]]_{5}$   & $[20,3,15]_5$& $g=(3 3 3 2 1 0 1 4 4 2 4 1 2 4 2 3 0 1)$ & $[[20,2,15/2]]_{5}$\\
\hline
\end{tabular}
\end{table*}
\begin{table*}
\caption{Optimal Pure Asymmetric CSS Codes over $\F_{7}$}
\label{table:F7Optimal}
\centering
\begin{tabular}{| c  l  c || c  l  c || c  l  c || c  l  c |}
\hline
\textbf{No.} & \textbf{AQC $Q$} & \textbf{Type}
& \textbf{No.} & \textbf{AQC $Q$} & \textbf{Type}
& \textbf{No.} & \textbf{AQC $Q$} & \textbf{Type}
& \textbf{No.} & \textbf{AQC $Q$} & \textbf{Type} \\
\hline
1 &  $[[9,5,3/2]]_{7} $ & GC & 39 &  $[[14,9,4/2]]_{7} $ & BC,CC & 77 &  $[[17,5,10/2]]_{7} $ & BC & 115 &  $[[21,15,3/3]]_{7} $ & CC\\
2 &  $[[9,4,4/2]]_{7} $ & BC & 40 &  $[[14,8,5/2]]_{7} $ & BC & 78 &  $[[17,4,11/2]]_{7} $ & BC & 116 &  $[[22,18,3/2]]_{7} $ & BC \\
3 &  $[[9,3,5/2]]_{7} $ & BC & 41 &  $[[14,7,6/2]]_{7} $ & BC & 79 &  $[[17,3,12/2]]_{7} $ & BC	& 117 &  $[[22,17,4/2]]_{7} $ & BC \\
4 &  $[[9,2,6/2]]_{7} $ & BC,GC & 42 &  $[[14,6,7/2]]_{7} $ & BC & 80 &  $[[17,2,13/2]]_{7} $ & BC & 118 &  $[[22,2,18/2]]_{7} $ & BC \\
5 &  $[[9,1,7/2]]_{7} $ & BC & 43 &  $[[14,4,8/2]]_{7} $ & BC & 81 &  $[[17,1,14/2]]_{7} $ & BC & 119 &  $[[23,19,3/2]]_{7} $ & BC \\

6 &  $[[10,6,3/2]]_{7} $ & BC & 44 &  $[[14,3,10/2]]_{7} $ & BC & 82 &  $[[18,14,3/2]]_{7} $ & BC	& 120 &  $[[23,18,4/2]]_{7} $ & BC \\
7 &  $[[10,5,4/2]]_{7} $ & BC & 45 &  $[[14,2,11/2]]_{7} $ & BC & 83 &  $[[18,13,4/2]]_{7} $ & BC	& 121 &  $[[24,20,3/2]]_{7} $ & BC,CC \\
\hline
\end{tabular}
\end{table*}
\begin{table*}
\caption{Nested Pairs of Cyclic Codes over $\F_{7}$ Yielding Optimal Asymmetric CSS Codes in Table~\ref{table:F7Optimal}}
\label{table:F7Cyclic}
\centering
\begin{tabular}{| l l l || l l l |}
\hline
\textbf{$C$ and $D$} & \textbf{Generator Polynomials} & \textbf{AQC $Q$} 
& \textbf{$C$ and $D$}& \textbf{Generator Polynomials} & \textbf{AQC $Q$}\\
\hline 
$[12,9,3]_7$&  $g=(2 4 4 1)$ & $[[12,6,3/3]]_7$ & $[19,10,8]_7$& $g=(6 5 2 0 3 1 3 5 6 1)$ & $[[19,7,8/3]]_7$ \\
$[12,3,6]_7$&  $(5 6 5 1 1 2 1)g$ & & $[19,3,15]_7$& $(6 6 6 6 6 2 2 1)g$ & \\

$[12,8,4]_7$&  $g=(1 6 2 6 1)$ & $[[12,5,4/3]]_7$ & $[19,16,3]_7$& $g=(6 3 3 1)$ & $[[19,6,9/3]]_7$\\
$[12,3,6]_7$&  $(4 1 2 6 2 1)g$ & & $[19,10,8]_7$& $(1 2 0 3 5 6 1)g$ &\\

$[12,8,4]_7$&  $g=(1 6 2 6 1)$ & $[[12,4,4/4]]_7$ & $[19,6,12]_7$& $g=(6 3 1 1 6 2 4 4 0 1 1 5 0 1)$ & $[[19,3,12/3]]_7$\\
$[12,4,6]_7$&  $(2 1 1 2 1)g$ & & $[19,3,15]_7$& $(6 3 3 1)g$ &\\

$[14,13,2]_7$& $g=(1 1)$ & $[[14,1,12/2]]_7$ & $[19,10,8]_7$& $g=(6 5 2 0 3 1 3 5 6 1)$ & $[[19,1,8/8]]_7$\\
\hline
\end{tabular}
\end{table*}
The list of best-known linear codes from the database of MAGMA yielding good AQCs 
will not be presented here since they can be searched and checked easily. For AQCs from group character 
codes, we simply enumerate them in Table~\ref{table:GC} according to the 
notations used in Theorem~\ref{thm:GCquantum}. 
\begin{table}
\caption{Nested Pairs of Group Character Codes Yielding Optimal Asymmetric CSS Codes 
in Tables~\ref{table:F3Optimal},~\ref{table:F5Optimal}, and~\ref{table:F7Optimal}}
\label{table:GC}
\centering
\begin{tabular}{| c | l | l |}
\hline
\textbf{$q$} & \textbf{$(r_{1},r_{2},l,t)$} & \textbf{AQC $Q$}\\
\hline
$3$ & $(1,2,3,2)$ & $[[8,3,4/2]]_{3}$ \\
    & $(1,3,4,2)$ & $[[16,10,4/2]]_{3}$ \\
    & $(1,4,5,2)$ & $[[32,25,4/2]]_{3}$ \\
$5$ & $(1,2,3,2)$ & $[[8,3,4/2]]_{5}$ \\
$7$ & $(1,3,2,3)$ & $[[9,5,3/2]]_7$ \\
    & $(2,3,2,3)$ & $[[9,2,6/2]]_7$ \\
\hline
\end{tabular}
\end{table}

\subsection{Tables of Good Pure Asymmetric CSS-like Codes for $q \in \{4,8,9\}$}\label{subsec:TabPowerPrimes}

In this subsection we list down good AQCs for $q \in \{4,8,9\}$ based on Table~\ref{table:good}. To qualify 
the goodness of a specific code under consideration, the theoretical LP bound for $k=\log_{q}K$ is 
explicitly provided. The defect \textbf{Def} is measured by subtracting the actual value of $k$ from the theoretical LP value.

Up to reasonable lengths, Tables~\ref{table:F4OptGood},~\ref{table:F8OptGood}, and~\ref{table:F9OptGood} 
contain good codes for $q \in \{4,8,9\}$. For brevity, in Table~\ref{table:F4OptGood} 
the convention that $\F_{4}=\F_{2}(w)$ where $w$ is a primitive root of an irreducible degree $2$ polynomial 
in $\F_{2}[x]$ is followed. 
Similarly, in Table~\ref{table:F8OptGood} we use the convention that $\F_{8}=\F_{2}(w)$ where $w$ is a primitive 
root of an irreducible polynomial of degree $3$ in $\F_{2}[x]$. In Table~\ref{table:F9OptGood} 
we let $\F_{9}=\F_{3}(w)$ where $w$ is a primitive root of a monic irreducible polynomial of 
degree $2$ in $\F_{3}[x]$. 

In Tables~\ref{table:F4AddCyclic},~\ref{table:F8AddCyclic}, and~\ref{table:F9AddCyclic}, 
nested subfield linear cyclic codes yielding good codes are listed down while 
Tables~\ref{table:F4Cyclic},~\ref{table:F8Cyclic}, and~\ref{table:F9Cyclic} present the cyclic pairs. 
We use the notations indicated in Theorem~\ref{thm:cyclic_main} and the abbreviation already mentioned 
above to write the generator polynomials.

\begin{table*}
\caption{good Pure Asymmetric CSS-like Codes over $\F_{4}$}
\label{table:F4OptGood}
\centering
\begin{tabular}{| c  l  c  c  c  c || c  l  c  c  c  c |}
\hline
 \textbf{No.} & \textbf{AQC $Q$} & \textbf{Type}
& \textbf{LP} & \textbf{Def} & \textbf{Remarks}
& \textbf{No.} & \textbf{AQC $Q$} & \textbf{Type}
& \textbf{LP} & \textbf{Def} & \textbf{Remarks} \\
\hline
1 & $[[6,2,4/2]]_{4}$ &	ACC & $2$ & $0$ & Optimal  & 69 & $[[15,6,5/3]]_{4}$ & CC &	$6.5$ & $0.5$& OpLin \\
2 & $[[6,1,3/3]]_{4}$ &	ACC & $1$ & $0$ & Optimal  & 70 & $[[15,5,6/3]]_{4}$ & CC & $5.5$ &	$0.5$& OpLin \\
3 & $[[7,3,3/2]]_{4}$ &	BC & $3.5$ & $0.5$ & OpLin & 71 & $[[15,4,7/3]]_{4}$ & CC & $4.5$ &	$0.5$& OpLin \\
4 & $[[7,2,4/2]]_{4}$ &	BC & $2.5$ & $0.5$ & OpLin & 72 & $[[15,1,10/3]]_{4}$ & CC & $1$ &$0$ & Optimal \\ 
5 & $[[7,1.5,5/2]]_{4}$ & ACC &	$1.5$ &	$0$& Optimal,BeOpLin & 73 & $[[15,5,5/4]]_{4}$ & CC  &$5.5$ & $0.5$& OpLin \\

6 & $[[8,4,3/2]]_{4}$ &	BC & $4.5$ &$0.5$ &	 OpLin & 74 & $[[15,4,6/4]]_{4}$ & CC  &$4.5$ &	$0.5$& OpLin \\
7 & $[[8,3,4/2]]_{4}$ &	BC & $3.5$ &$0.5$ &	 OpLin & 75 & $[[16,12,3/2]]_{4}$ & BC &$12$  &	$0$  & Optimal \\
\hline
\end{tabular}
\end{table*}

\begin{table*}
\caption{Nested Pairs of $\F_{2}$-linear Cyclic Codes over $\F_{4}$ Yielding Optimal or Good Asymmetric CSS-like Codes in Table~\ref{table:F4OptGood}}
\label{table:F4AddCyclic}
\centering
\begin{tabular}{| l l l |}
\hline
 \textbf{$C$ and $D$} & \textbf{Generator Polynomials} & \textbf{AQC $Q$} \\
\hline 
$(6,2^{6},4)_4$ & $g_{4}=(1011)+w(11)$, $g_{2}=(111111)$ &$[[6,2,4/2]]_4$\\
$(6,2^{2},6)_4$ & $(10101)g_{4}$, $(1)g_{2}$ &\\

$(6,2^{7},3)_4$ & $g_{4}=(101)+w(11)$, $g_{2}=(10101)$ &$[[6,1,3/3]]_4$\\
$(6,2^{5},4)_4$ & $(11)g_{4}$, $(11)g_{2}$      &\\

$(7,2^{5},5)_4$ & $g_{4}=(11101)+w(1011)$, $g_{2}=(1111111)$ &$[[7,1.5,5/2]]_4$\\
$(7,2^{2},7)_4$ & $(1101) g_{4}$, $(1) g_{2}$      &\\

$(12,2^{13},5)_4$ & $g_{4}=(001110001)+w(1)$, $g_{2}=(111111111111)$ &$[[12,5.5,5/2]]_4$\\
\hline
\end{tabular}
\end{table*}
\begin{table*}
\caption{Nested Pairs of Linear Cyclic Codes over $\F_{4}$ Yielding Optimal or Good Asymmetric CSS Codes in Table~\ref{table:F4OptGood}}
\label{table:F4Cyclic}
\centering
\begin{tabular}{| l l l || l l l |}
\hline
\textbf{$C$ and $D$} & \textbf{Generator Polynomials} & \textbf{AQC $Q$} 
& \textbf{$C$ and $D$}& \textbf{Generator Polynomials} & \textbf{AQC $Q$}\\
\hline
$[11,6,5]_4$    &$g=(1 w^211   w1)$     &$[[11,1,5/5]]_4$ & $[15,12,3]_4$   &$g=(w^2011)$     &$[[15,1,10/3]]_4$\\
$[11,5,6]_4$    &$(11)g$      & & $[15,11,4]_4$   &$(  w1)g$      &\\

$[15,12,3]_4$   &$g=(w^2011)$     &$[[15,9,3/3]]_4$ & $[15,9,5]_4$    &$g=(1   w11 w^2 w^21)$ &$[[15,5,5/4]]_4$\\
$[15,3,11]_4$   &$(w^20   w00 w^210   w1)g$ & & $[15,4,10]_4$   &$(w^2   w0 w^201)g$      &\\

$[15,12,3]_4$   &$g=(w^2011)$     &$[[15,8,4/3]]_4$ & $[15,8,6]_4$    &$g=(  w w^2   w0   w0 w^21)$&$[[15,4,6/4]]_4$\\
$[15,4,10]_4$   &$(w^2 w^2111   w w^2 w^21)g$  & & $[15,4,10]_4$   &$(  w1001)g$  &\\

$[15,12,3]_4$   &$g=(w^2011)$     &$[[15,6,5/3]]_4$ & $[17,13,4]_4$   &$g=(11   w11)$ &$[[17,9,4/4]]_4$\\
$[15,6,8]_4$    &$(  w   w   w0 w^201)g$  & & $[17,4,12]_4$   &$(1001111001)g$ &\\

$[15,8,6]_4$    &$g=(  w w^2   w0   w0 w^21)$&$[[15,5,6/3]]_4$ & $[17,9,7]_4$    &$g=(1 w^20 w^2 w^2 w^2 0 w^21)$&$[[17,5,7/4]]_4$\\
$[15,3,11]_4$   &$(  w w^20   w01)g$      & & $[17,4,12]_4$   &$(1   w   w   w   w1)g$      &\\

$[15,12,3]_4$   &$g=(w^2011)$     &$[[15,4,7/3]]_4$ & $[17,8,8]_4$    &$g=(1   w w^2 w^200 w^2 w^2w1)$&$[[17,4,8/4]]_4$\\
$[15,8,6]_4$    &$(1 w^2 w^2 w^21)g$  & & $[17,4,12]_4$   &$(1 w^21 w^21)g$  &\\
\hline
\end{tabular}
\end{table*}
\begin{table*}
\caption{Good Pure Asymmetric CSS-like Codes over $\F_{8}$}
\label{table:F8OptGood}
\centering
\begin{tabular}{| c  l  c  c  c  c || c l c c c c |}
\hline
 \textbf{No.} & \textbf{AQC $Q$} & \textbf{Type}
& \textbf{LP} & \textbf{Def} & \textbf{Remarks}
& \textbf{No.} & \textbf{AQC $Q$} & \textbf{Type}
& \textbf{LP} & \textbf{Def} & \textbf{Remarks} \\
\hline
1  & $[[10,19/3,3/2]]_{8}$ & ACC& $20/3$ & $1/3$ & BeOpLin & 44  & $[[15,7,7/2]]_{8}$ & BC &	$23/3$ & $2/3$	& OpLin \\
2  & $[[10,6,4/2]]_{8}$    & BC & $6$ 	 & $0$	 & Optimal & 45  & $[[15,6,8/2]]_{8}$ & BC &$20/3$ & $2/3$	& OpLin \\
3  & $[[10,4,5/2]]_{8}$    & BC & $14/3$ & $2/3$ & OpLin & 46  & $[[15,4,10/2]]_{8}$ & BC &$13/3$ & $1/3$	& OpLin \\
4  & $[[10,3,6/2]]_{8}$    & BC & $11/3$ & $2/3$ & OpLin & 47  & $[[15,3,11/2]]_{8}$ & BC &$10/3$ & $1/3$	& OpLin \\
5  & $[[10,2,8/2]]_{8}$    & BC & $2$	 & $0$	 & Optimal & 48  & $[[15,2,12/2]]_{8}$ & BC &$7/3$ & $1/3$	& OpLin \\ 

6  & $[[11,7,3/2]]_{8}$ & BC &	$23/3$	& $2/3$	& OpLin & 49  & $[[16,12,3/2]]_{8}$ & BC &$38/3$ & $2/3$	& OpLin \\
7  & $[[11,6,4/2]]_{8}$ & BC &	$20/3$	& $2/3$	& OpLin & 50  & $[[16,11,4/2]]_{8}$ & BC &$35/3$ & $2/3$	& OpLin \\
\hline
\end{tabular}
\end{table*}
\begin{table*}
\caption{Nested Pairs of $\F_{2}$-Linear Cyclic Codes over $\F_{8}=\F_{2}(w)$ Yielding Optimal or Good Asymmetric CSS-like Codes in Table~\ref{table:F8OptGood}}
\label{table:F8AddCyclic}
\centering
\begin{tabular}{| l l l |}
\hline
 \textbf{$C$ and $D$} & \textbf{Generator Polynomials} & \textbf{AQC $Q$}\\
\hline
$(10,2^{22},3)_8$    &$g_{1} =(0 00011)+w(0)+w^2(1)$, $g_{2} =(11 01001)+w(1)$, $g_{3} =(101010101)$ & $[[10,19/3,3/2]]_8$\\
$(10,2^{3},10)_8$    &$(1111111111)g_{1} $, $(1111111111)g_{2} $, $(11)g_{3} $ &\\

$(15,2^{37},3)_8$ &$g_{1} =(1001)+w(0)+w^2(111)$, $g_{2} =(010011)+w(1)$, $g_{3} =(1001111)$ &$[[15,34/3,3/2]]_8$\\
$(15,2^{3},15)_8$       &$(1001001001001)g_{1} $, $(11011 011 011011)g_{2} $, $(1001110011)g_{3} $ & \\

$(15,2^{34},4)_8$   &$g_{1} =(111011111)+w( 0)+w^2(111)$, $g_{2} =(0 0 00111)+w(1)$, $g_{3} =(1001110011)$ &$[[15,31/3,4/2]]_8$\\
$(15,2^{3},15)_8$   &$(1001001001001)(111)g_{1} $, $(11 011 011011011)g_{2} +(1001001001001)(011)g_{1} $, $(1001111)g_{3} $ &\\

$(17,2^{43},3)_8$       &$g_{1} =(0 0100101)+w(0)+w^2(1)$, $g_{2} =(0100 0101)+w(1)$, $g_{3} =(111010111)$ &$[[17,40/3,3/2]]_8$\\
$(17,2^{3},17)_8$       &$(11111111111111111)g_{1} $, $(11111111111111111)g_{2} $, $(10011100 1)g_{3} $ &\\
\hline
\end{tabular}
\end{table*}
\begin{table*}
\caption{Nested Pairs of Linear Cyclic Codes over $\F_{8}$ Yielding Optimal or Good Asymmetric CSS Codes in Table~\ref{table:F8OptGood}}
\label{table:F8Cyclic}
\centering
\setlength{\tabcolsep}{3pt}
\begin{tabular}{| l l l || l l l |}
\hline
 \textbf{$C$ and $D$} & \textbf{Generator Polynomials} & \textbf{AQC $Q$} 
& \textbf{$C$ and $D$}& \textbf{Generator Polynomials} & \textbf{AQC $Q$}\\
\hline
$[13,9,4]_8$    &$ g=(1 w^3 w^5 w^3 1)$ &$[[13,8,4/2]]_8$ & $[21,20,2]_8$   &$ g=(1 1)$     &$[[21,17,3/2]]_8$\\
$[13,1,13]_8$   &$(1 w 0 w^5 w^2 w^5 0 w 1) g$  & & $[21,3,14]_8$   &$(w^5 w^6 w^4 w^2 1 w^3 1 w^6 w^3 w^2 0 w^4 w^5 w 0 0 w 1) g$ & \\

$[14,11,3]_8$   &$ g=(w^2 w^3 w^6 1)$     &$[[14,10,3/2]]_8$ & $[21,20,2]_8$   &$ g=(11)$     &$[[21,16,4/2]]_8$\\
$[14,1,14]_8$   &$(w^5 w w^5 w^4 1 w^2 w^4 w 0 w^2 1) g$ & & $[21,4,14]_8$   &$(w^5 0 w^3 w w^5 w^3 w^3 w^4 w^5 w^2 w^4 w^4 w w^6 w w^3 1) g$ &\\

$[14,10,4]_8$   &$ g=(w w^4 1 w^2 1)$ &$[[14,9,4/2]]_8$ & $[21,18,3]_8$   &$ g=(1 1 0 w^3 1)$ &$[[21,15,3/3]]_8$\\
$[14,1,14]_8$   &$(w w^6 w w^2 w 1 w^6 w^6 w^3 1) g$ & & $[21,3,14]_8$   &$(w^5 w^6 w^2 w^6 w 0 1 w^5 w w^4 w^3 w^4 w^2 w^3 0 1) g$ &\\

$[14,2,12]_8$   &$ g=(w^6 w^5 1 w^3 1 1 0 w^6 w^5 1 w^3 1 1)$&$[[14,1,12/2]]_8$ & $[21,17,4]_8$   &$ g=(1 w w^3 w 1)$ &$[[21,14,4/3]]_8$\\
$[14,1,14]_8$   &$(w^6 1) g$      & & $[21,3,14]_8$   &$(1 1 w^6 w^6 w^6 1 w^3 w^6 1 w w^4 w w^5 w^3 1) g$  &\\

$[19,13,6]_8$   &$ g=(1 w^3 w^6 w^6 w^6 w^3 1)$ &$[[19,12,6/2]]_8$ & $[21,17,4]_8$   &$ g=(1 w w^3 w 1)$ &$[[21,13,4/4]]_8$\\
$[19,1,19]_8$   &$(1 w w w^3 1 w^6 w w^6 1 w^3 w w 1) g$ & & $[21,4,14]_8$   &$(w^6 w^4 w w^6 w^3 w^2 w^6 w^5 w^6 w^5 w^6 w^6 1 1) g$ &\\
\hline
\end{tabular}
\end{table*}

\begin{table*}
\caption{good Pure Asymmetric CSS-like Codes over $\F_{9}$}
\label{table:F9OptGood}
\centering
\begin{tabular}{| c  l  c  c  c  c || c  l  c  c  c  c |}
\hline
 \textbf{No.} & \textbf{AQC $Q$} & \textbf{Type}
& \textbf{LP} & \textbf{Def} & \textbf{Remarks}
& \textbf{No.} & \textbf{AQC $Q$} & \textbf{Type}
& \textbf{LP} & \textbf{Def} & \textbf{Remarks} \\
\hline
1 & $[[10,7,3/2]]_9 $ &	BC & $7$ & $0$ & Optimal & 50 & $[[15,9,5/2]]_9 $ & BC  &	$9.5$ & $0.5$ & OpLin  \\
2 & $[[10,6,4/2]]_9 $ &	CC & $6$ & $0$ & Optimal & 51 & $[[15,8,6/2]]_9 $ & BC  &	$8.5$ & $0.5$ & OpLin  \\
3 & $[[10,5,5/2]]_9 $ &	BC & $5$ & $0$ & Optimal & 52 & $[[15,7,7/2]]_9 $ & BC  &	$7.5$ & $0.5$ & OpLin  \\
4 & $[[10,4,6/2]]_9 $ &	CC & $4$ & $0$ & Optimal & 53 & $[[15,6,8/2]]_9 $ & BC  &	$6.5$ & $0.5$ & OpLin  \\
5 & $[[10,3,7/2]]_9 $ &	BC & $3$ & $0$ & Optimal & 54 & $[[15,5,9/2]]_9 $ & BC  &	$5.5$ & $0.5$ & OpLin  \\

6 & $[[10,2,8/2]]_9 $   & CC & $2$   & $0$	 & Optimal& 55 & $[[15,4,10/2]]_9 $ & BC &	$4.5$ &	$0.5$ & OpLin  \\
7 & $[[10,2,6/4]]_9 $   & CC  &	$2$  & $0$   & Optimal & 56 & $[[15,3,11/2]]_9 $ & BC &	$3.5$ &	$0.5$ & OpLin  \\
\hline
\end{tabular}
\end{table*}
\begin{table*}
\caption{Nested Pairs of $\F_{3}$-Linear Cyclic Codes over $\F_{9}=\F_{3}(w)$ Yielding Optimal or Good Asymmetric CSS-like Codes in Table~\ref{table:F9OptGood}}
\label{table:F9AddCyclic}
\centering
\begin{tabular}{| l l l |}
\hline
\textbf{$C$ and $D$} & \textbf{Generator Polynomials} & \textbf{AQC $Q$}\\
\hline
$(11,3^{17},3)_9$ & $g_{9}=(0 1 1)+w(1)$, $g_{3}=(2 0 1 2 1 1)$ &$[[11,7.5,3/2]]_9$\\
$(11,3^{2},11)_9$       & $(1 1 1 1 1 1 1 1 1 1 1)g_{9}$, $(2 2 1 2 0 1)g_{3}$ &\\

$(11,3^{17},3)_9$ &$g_{9}=(0 1 1)+w(1)$, $g_{3}=(2 0 1 2 1 1)$ &$[[11,6,3/3]]_9$\\
$(11,3^{5},8)_9$        & $(1 0 1 2 2 2 1)g_{9}$, $(1 0 1 2 2 2 1)g_{3}$ &\\

$(11,3^{11},6)_9$ &$g_{9}=(1 0 1 0 2 1 1)+w(2 1)$, $g_{3}=(1 1 1 1 1 1 1 1 1 1 1)$&$[[11,3,6/3]]_9$\\
$(11,3^{5},8)_9$        & $(2 0 1 2 1 1)g_{9}$, $(2 1)g_{3}$  &\\

$(12,3^{19},3)_9$ &$g_{9}=(0 0 1)+w(1 1)$, $g_{3}=(1 2 2 2 1)$ &$[[12,8.5,3/2]]_9$\\
\hline
\end{tabular}
\end{table*}

\begin{table*}
\caption{Nested Pairs of Linear Cyclic Codes over $\F_{9}$ Yielding Optimal or Good Asymmetric CSS Codes in Table~\ref{table:F9OptGood}}
\label{table:F9Cyclic}
\centering
\begin{tabular}{| l l l || l l l |}
\hline
 \textbf{$C$ and $D$} & \textbf{Generator Polynomials} & \textbf{AQC $Q$} 
& \textbf{$C$ and $D$}& \textbf{Generator Polynomials} & \textbf{AQC $Q$}\\
\hline

$[10,7,4]_9$    &$g=(1 w^2 w^21)$     &$[[10,6,4/2]]_9$ & $[14,10,4]_9$   &$g=(2 w^2 w^2 w^2 1)$ &$[[14,9,4/2]]_9$\\
$[10,1,10]_9$   &$(1 w^5 w^2 w^6 w^2 w^5 1)g$  & & $[14,1,14]_9$   &$(1 w 0 1 1 1 1 0 w^3 1)g$ &\\

$[10,9,2]_9$    &$g=(2 1)$     &$[[10,4,6/2]]_9$ & $[16,13,3]_9$   &$g=(1 w^3 w^5 1)$     &$[[16,12,3/2]]_9$\\
$[10,5,6]_9$    &$(1 w^2 0 w^2 1)g$  & & $[16,1,16]_9$   &$(1 w w^7 0 w^6 w^7 2 1 w^3 w^5 1 w^2 1)g$ &\\

$[10,9,2]_9$    &$g=(21)$     &$[[10,2,8/2]]_9$ & $[16,12,4]_9$   &$g=(w^6 w^5 0 2 1)$ &$[[16,11,4/2]]_9$\\
$[10,7,4]_9$    &$(1 w^5 1)g$  & & $[16,1,16]_9$   &$(w 0 w^7 w^3 w^3 1 w^6 w^3 w^5 w^6 w^2 1)g$ &\\

$[10,7,4]_9$    &$g=(1 w^2 w^2 1)$     &$[[10,2,6/4]]_9$ & $[16,2,14]_9$   &$g=(w^3 w^3 w 2 1 w 1 0 w^3 w^3 w 2 1 w 1)$ &$[[16,1,14/2]]_9$\\
$[10,5,6]_9$    &$(1 w^7 1)g$  & & $[16,1,16]_9$   &$(w^2 1)g$      &\\

$[11,6,5]_9$    &$g=(2 0 1 2 1 1)$     &$[[11,5,5/2]]_9$ & $[20,17,3]_9$   &$g=(1 w^2 w^5 1)$     &$[[20,16,3/2]]_9$\\
$[11,1,11]_9$   &$(2 2 1 2 0 1)g$      & & $[20,1,20]_9$   &$(w^2 0 w 2 0 w^7 w w^2 w^2 w^2 0 w^2 1 w^7 w^2 2 1)g$   &\\

$[13,10,3]_9$   &$g=(2 1 1 1)$     &$[[13,9,3/2]]_9$ & $[20,16,4]_9$   &$g=(w^6 2 w^7 0 1)$ &$[[20,15,4/2]]_9$\\
$[13,1,13]_9$   &$(2 1 2 1 0 2 2 0 0 1)g$& & $[20,1,20]_9$   &$(2 w^2 w w^2 1 w w^3 w^3 w^7 w^7 w^5 2 w^6 w^5 w^6 1)g$       &\\
\hline
\end{tabular}
\end{table*}

\subsection{Some \textit{Ad Hoc} Constructions}\label{subsec:AdHoc}
In some cases, an \textit{ad hoc} construction of suitable nested pairs of 
classical codes indicated by the linear programming output yields AQCs with optimal $k$.

We show an explicit construction for codes $Q$
with parameters $[[4,1,2/2]]_{2}$ and $[[5,2,2/2]]_{2}$ by exhibiting a 
generator matrix for each of $[4,2,2]_{2}$ and $[5,3,2]_{2}$-codes containing 
a codeword of weight $4$ and $5$, respectively. The matrices are
\begin{equation*}
\left(
\begin{array}{c}
1\,0\,1\,0 \\
0\,1\,0\,1 \\
\end{array}
\right) \text{ and }
\left(
\begin{array}{c}
1\, 0\, 0\, 0\, 1 \\
0\, 1\, 0\, 1\, 1 \\
0\, 0\, 1\, 0\, 1 \\
\end{array}
\right) \text{.}
\end{equation*} 

A $[[13,1,5/3]]_{2}$-code alluded to by a referee of~\cite{SKR09} 
can also be constructed as shown below. It is rather surprising that this CSS code is optimal, 
given that it is far from reaching the quantum Singleton bound. It is of interest to know 
if the construction here is essentially the only one possible, up to code equivalence.

Consider the best-known $[13,9,3]_{2}$-linear code $C_{2}$ from the MAGMA database. 
Its dual $C_{2}^{\perp_{\E}}$ is a $[13,4,6]_{2}$-code. Next, we add the row vector 
$\1:=(1,1,\ldots,1)$ to the generator matrix of $C_{2}^{\perp_{\E}}$ to
get a $[13,5,5]_{2}$-code $C_{1}$. The dual of $C_{1}$ is a $[13,8,4]_{2}$-code which
is a subcode of $C_{2}$. Hence, we can use $C_{1}^{\perp_{\E}} \subset C_{2}$ to get 
a $[[13,1,5/3]]_{2}$-quantum code $Q$.

Recently, some new best-known linear codes over $\F_{5}$ are presented in~\cite[Sec. 5]{JLLX10}. 
The first one is of parameters $[36,28,6]_{5}$ and is labeled $\mathcal{C}_{36}$ in the 
said reference.  By shortening this code at the first position a $[35,27,6]_{5}$-code 
is derived. If we shorten $\mathcal{C}_{36}$ at the first two positions, 
we get a $[34,26,6]_{5}$-code. It can be easily checked, starting with the generator matrix 
of $\mathcal{C}_{36}$, that each of these three codes contains a codeword $\v$ of weight 
equal to its length. By Theorem~\ref{thm:spectrumn}, we get CSS codes with 
parameters $[[36,27,6/2]]_{5}, [[35,26,6/2]]_{5}$, and $[[34,25,6/2]]_{5}$, all of which are optimal.

\section{Conclusion and Open Problems}\label{sec:conclude}

It is instructive to consider, by way of simple examples presented in Table~\ref{table:comp}, 
the difference between the best performing symmetric quantum codes and their asymmetric counterparts. 
AQCs allow us to tailor the process of error-correction better once the ratio of asymmetry 
in the channel is known or can be approximated properly. Extensive data on the best-known symmetric 
quantum codes, given $n$ and $k$ for $q=2$, can be found in~\cite{Gr09}.
\begin{table}
\caption{Some Examples Comparing Symmetric and Asymmetric quantum codes for $q=2$}
\label{table:comp}
\centering
\begin{tabular}{| c | c | c | c |}
\hline
 \textbf{$n$} & \textbf{$k$} & \textbf{Symmetric} & \textbf{Asymmetric}\\
             &               & \textbf{$d$}       & \textbf{$(d_{z},d_{x})$} \\
\hline
$7$ & $3$ & $2$ & $(3,2)$ \\
$8$ & $3$ & $3$ & $(4,2)$ \\
$15$ & $1$ & $5$ & $(7,3)$ \\
\hline
\end{tabular}
\end{table}

In this paper, the functional approach is used to establish CSS-like constructions allowing us 
to use pairs of nested subfield linear codes over $\F_{q}$ to construct pure AQCs. The standard 
CSS construction is shown to be a special case.

Combining specific constructions of pairs of nested classical codes and linear programming, 
we show that CSS-like constructions based on pairs of subfield linear codes over $\F_{q}$ 
often give us optimal or good $q$-ary pure AQCs with better parameters than 
the best that the standard CSS construction can achieve. Lists of optimal or best known 
pure CSS-like AQCs up to some computationally reasonable length for $q \in \{2,3,4,5,7,8,9\}$ 
are given in the hope of providing a more comprehensive list of best performing AQCs.

While working on earlier versions of this paper, we found out that the linear programming (LP) 
approach was sufficiently effective for small values of $q$ and $n$. At the same time, the loss of 
precision due to the extremely large coefficients involved in the computation soon became very limiting as 
these values grew larger, as long as traditional LP solvers such as CPLEX were used.

To handle larger values of $q$ and $n$, we started experimenting with arbitrary precision LP solvers, such as 
PPL~\cite{ppl}, which is now equipped with Sage~\cite{SAGE} interface, thanks largely to the efforts of Risan, 
then an undergraduate student at Nanyang Technological University, and the last author as presented in~\cite{Pas12} 
and in~\cite{RP12}.

The initial results in this direction were extremely encouraging, and we are able to solve most, if not all, of the 
LP instances we have previously encountered as intractable by traditional LP solvers.   
More efforts still need to be exerted in perfecting our software, in coming up with better upper bounds, and 
in constructing AQCs meeting the bounds.

The stabilizer formalism of symmetric quantum codes can be extended naturally to the asymmetric case. 
How the CSS-like constructions are connected to stabilizer AQCs is an interesting question to explore. 

\section*{Appendix A: Proof of Theorem~\ref{thm:closure}}
Given that $\F_p \subseteq \F_{r} \subseteq \F_{q}$, we equip the space
$\F_{q}^{n}$ with the trace Euclidean inner product.

\renewcommand{\thetheorem}{A.1}
\begin{lemma}
$\langle \u, \v \rangle_{\TrE}$ is a valid inner product on $\F_{q}^{n}$.
\end{lemma}
\begin{IEEEproof}
The only property to check is non-degeneracy since everything else follows
immediately from $\langle \u, \v \rangle_{\E}$. We show that if
$\langle \u, \v \rangle_{\TrE} = 0$ for all $\u \in \F_{q}^{n}$,
then $\v = \0$, the converse being trivial.

Let us assume that $\v \neq \0$ and construct a vector $\u \in \F_{q}^{n}$ such that 
$\langle \u, \v \rangle_{\TrE} \neq 0$ to settle the claim. Since the trace mapping is onto, 
there exists $0 \neq a \in \F_{q}$ such that $\Trq(a) \neq 0$. Let $j$ be 
the first index for which $v_{j} \neq 0$. Define $\u \in \F_{q}^n$ as follows: 
$u_{j}=v_{j}^{-1}a$ and $u_{i}=0$ for all $i \neq j$.
\end{IEEEproof}

\begin{IEEEproof}[Proof of Theorem~\ref{thm:closure}]
Let $A(Y):=\sum_{0}^{n} A_{j}Y^{j}$ and $B(Y):=\sum_{0}^{n} B_{j}Y^{j}$ be,
respectively, the weight enumerators of $C$ and of $\CduTrE$.
We first prove the following identity
\begin{equation}\label{eq:sizedual}
B(Y) = \frac {\left(1+(q-1)Y\right)^{n}}{|C|} \cdot A\left(\frac{1-Y}{1+(q-1)Y}\right)\text{.}
\end{equation}

Let $\c=(c_{1},\ldots,c_{n}),\d=(d_{1},\ldots,d_{n}) \in C$. Let $\chi$ be a nontrivial additive character of $\F_{r}$. Since $C$ is $\F_{r}$-linear, we 
can define for every $\b=(b_{1},\ldots,b_{n}) \in \F_{q}^{n}$ a character $\chi_{\b}$ of the additive group $C$ by 
substituting the trace Euclidean form for the argument of the character $\chi$, such that
\begin{align*}
 \chi_{\b}(\c) & = \chi \left( \bdotcTrE \right) \\
               & = \chi \left( \sum_{i=1}^{n} \Trq (b_{i} c_{i}) \right) \in \C \text{.}
\end{align*}
The character $\chi_{\b}$ is trivial if and only if $\b \in \CduTrE$. 
Thus, we have the orthogonality relation of characters
\begin{equation}\label{eq:orthochar}
 \sum_{\c \in C} \chi_{\b} (\c) =
\begin{cases}
|C| &\text{ if } \b \in \CduTrE \\
0 &\text{ otherwise}
\end{cases}
\text{.}
\end{equation}

By (\ref{eq:orthochar}),
\begin{align}\label{eq:2.4}
 \sum_{\c \in C} \sum_{\b \in \F_{q}^{n}} \chi_{\b}(\c) Y^{\wt (\b)}
&= \sum_{\b \in \F_{q}^{n}} Y^{\wt (\b)} \sum_{\c \in C} \chi_{\b} (\c) \notag\\
&= |C| \sum_{i=0}^{n} B_{i} Y^{i} = |C|\cdot B(Y) \text{.}
\end{align}

Let us take a closer look at the inner sum on the left hand side of
(\ref{eq:2.4}). By the property of the trace mapping, we can distribute
the trace mapping over each coordinate.
\begin{align}\label{eq:2.5}
 & \sum_{\b \in \F_{q}^{n}} \chi_{\b} (\c) Y^{\wt (\b)} \notag \\
 & = \sum_{\left( b_{1},b_{2}, \ldots, b_{n}\right) \in \F_{q}^{n}}
    \left(Y^{ \left(\sum\limits_{i=1}^{n} \wt(b_{i}) \right)} \right)
    \chi \left( \sum_{i=1}^{n} \Trq (b_{i} c_{i})\right) \notag \\
 & = \sum_{\left( b_{1},b_{2}, \ldots, b_{n}\right) \in \F_{q}^{n}}
    \left( \prod\limits_{i=1}^{n} Y^{\wt(b_{i})} \cdot \chi \left(\Trq(b_{i} c_{i}) \right) \right) \notag \\
 & = \prod_{i=1}^{n} \sum_{b_{i} \in \F_{q}} Y^{\wt(b_{i})} \chi \left( \Trq (b_{i} c_{i})\right) \text{.}
\end{align}
Note that if $(c_{i}, b_{i})=(0,0)$, the contribution to the sum in the right hand side is $1$.
If $c_{i}=0$ but $b_{i} \neq 0$, the contribution to the sum is $(q-1)Y$.
Similarly, for $c_{i} \neq 0$, if $b_{i}=0$, we get $1$, while if $b_{i} \neq 0$,
we get $-Y$. Therefore, the sum in the right hand side of (\ref{eq:2.5}) can be simplified to
\begin{equation}
 \sum_{b_{i} \in \F_{q}} Y^{\wt(b_{i})} \chi \left( \Trq (b_{i} c_{i})\right) =
 \begin{cases}
1+(q-1)Y &\text{ if } c_{i}=0 \\
1-Y      &\text{ if } c_{i} \neq 0
 \end{cases}\text{,}
\end{equation}
yielding, after plugging this result back to (\ref{eq:2.4}),
\begin{align*}
 B(Y) &= \frac{1}{|C|} \sum_{\c \in C} \sum_{\b \in \F_{q}^{n}} \chi_{\b} (\c) Y^{\wt(\b)} \\
      &= \frac{1}{|C|} \sum_{\c \in C} \left(1+(q-1)Y \right)^{n-\wt(\c)} (1-Y)^{\wt(\c)} \\
      &= \frac{1}{|C|} \left(1+(q-1)Y \right)^{n} \sum_{\c \in C} \left( \frac{(1-Y)}{1+(q-1)Y} \right)^{\wt(\c)} \\
      &= \frac{(1+(q-1)Y)^{n}}{|C|} \cdot A\left( \frac{(1-Y)}{1+(q-1)Y} \right) \text{.}
\end{align*}
This establishes (\ref{eq:sizedual}).

Hence, by using the definition of $A(Y)$, $B(Y)$ can now be written as
\begin{equation*}
B(Y) = \frac{1}{|C|} \sum_{i=0}^{n} A_{i}(1+(q-1)Y)^{n-i}(1-Y)^{i} \text{.}
\end{equation*}
Comparing the coefficients of $Y^{j}$ on both sides gives us the claimed MacWilliams
equation for the single variable $Y$. Replacing $Y$ by $\frac{Y}{X}$ and multiplying 
both sides by $X^{n}$ give the desired expression for the two-variable case.

Note that $|\CduTrE|$ can be derived by substituting $Y=1$ in 
(\ref{eq:sizedual}) from whence we have $|C||\CduTrE| = q^{n}$. This is sufficient 
to establish the closure property of $C$ under the trace Euclidean inner product.
\end{IEEEproof}

\renewcommand{\thetheorem}{A.2}
\begin{remark}
The closure property and the MacWilliams equation for an $\F_{r}$-linear code $C$
over $\F_{q}$ under the trace Euclidean inner product can also be deduced
from~\cite[Cor. 3.2.3 on p. 88]{NRS06}. The explicit approach above is preferred
so as to eliminate the need for a more sophisticated algebraic build-up in the exposition.
\end{remark}

\section*{Appendix B: Proof of Theorem~\ref{thm:TrEandTrH}}
We begin with some preparatory lemmas. Recall that for $C \subseteq \F_{q}^n$, 
$\overline{C}:=\{ \overline{\c} : \c \in C \}$.

\renewcommand{\thetheorem}{B.1}
\begin{lemma}\label{lem7}
Suppose that $q=r^2$ is odd. Let $C_{1}$ and $C_{2}$ be $\F_{r}$-linear
codes of length $n$ over $\F_{q}$. For $\alpha\in \F_{q} \setminus \{0\}$
such that $\overline{\alpha}=-\alpha$, the following statements hold:
\begin{enumerate}[$i)$]
\item If $C_{1}^{\perp_{\TrE}} \subseteq C_{2}$, then
$$\alpha^{-1}\overline{C_{1}^{\perp_{\TrE}}} \subseteq
\left( C_{2}^{\perp_{\TrE}} \right)^{\perp_{\Tr\H}}\text{.}$$
\item If $C_{1}^{\perp_{\Tr\H}} \subseteq C_{2}$, then
$$\alpha \overline{C_{1}^{\perp_{\Tr\H}}} \subseteq
\left( C_{2}^{\perp_{\Tr\H}} \right)^{\perp_{\TrE}}\text{.}$$
\end{enumerate}
\end{lemma}

\begin{IEEEproof}
Let $\u \in C_{1}^{\perp_{\TrE}}$ and $\v \in C_{2}^{\perp_{\TrE}}$. Then,
\begin{equation*}
 0 = \langle \u, \v \rangle_{\TrE} =
\langle \u,\v \rangle_{\E} + \overline{ \langle \u,\v \rangle_{\E}}\text{.}
\end{equation*}
Therefore,
\begin{align*}
\langle \alpha^{-1}\overline{\u}, \v \rangle_{\Tr\H} & =
\alpha \langle \alpha^{-1} \overline{\u},\v \rangle_{\rm H} +
\overline{\alpha \langle \alpha^{-1} \overline{\u},\v \rangle_{\rm H}}\\
& = \overline {\langle \u, \v \rangle_{\E}} + \langle \u, \v \rangle_{\E} = 0 \text{.}
\end{align*}
Hence, $\alpha^{-1}\overline{C_{1}^{\perp_{\TrE}}} \subseteq
\left( C_{2}^{\perp_{\TrE}} \right)^{\perp_{\Tr\H}}$. This proves $i)$.

To prove $ii)$, let $\u \in C_{1}^{\perp_{\Tr\H}}$ and
$\v \in C_{2}^{\perp_{\Tr\H}}$. Then
\begin{equation*}
 0 = \langle \u,\v \rangle_{\Tr\H} = \alpha \langle \u,\overline{\v} \rangle_{\E}
+ \overline{\alpha} \langle \overline{\u},\v \rangle_{\E}\text{.}
\end{equation*}

Since $\overline{\alpha} = - \alpha$,
\begin{align*}
\langle \alpha \overline{\u}, \v \rangle_{\TrE} & =
\langle \alpha \overline{\u}, \v \rangle_{\E} +
\langle \overline{\alpha} \u, \overline{\v} \rangle_{\E} \\
& = \alpha \langle \overline{\u}, \v \rangle_{\E} +
\overline{\alpha} \langle \u, \overline{\v} \rangle_{\E} \\
& = - \left( \overline{\alpha} \langle \overline{\u}, \v \rangle_{\E}
+ \alpha \langle \u, \overline{\v} \rangle_{\E} \right)= 0 \text{.}
\end{align*}
Therefore, $\alpha \overline{C_{1}^{\perp_{\Tr\H}}} \subseteq
\left( C_{2}^{\perp_{\Tr\H}} \right)^{\perp_{\TrE}}$.
\end{IEEEproof}

\renewcommand{\thetheorem}{B.2}
\begin{lemma}\label{lem8}
Suppose that $q=r^2$ is even. Let $C_{1}$ and $C_{2}$ be $\F_{r}$-linear
codes of length $n$ over $\F_{q}$. Then the following statements hold:
\begin{enumerate}[$i)$]
\item If $C_{1}^{\perp_{\TrE}} \subseteq C_{2}$, then
$$\overline{C_{1}^{\perp_{\TrE}}} \subseteq
\left( C_{2}^{\perp_{\TrE}} \right)^{\perp_{\Tr\H}}.$$
\item If $C_{1}^{\perp_{\Tr\H}} \subseteq C_{2}$, then
$$\overline{C_{1}^{\perp_{\Tr\H}}} \subseteq
\left( C_{2}^{\perp_{\Tr\H}} \right)^{\perp_{\TrE}}.$$
\end{enumerate}
\end{lemma}
\begin{IEEEproof}
The proof follows from the proof of Lemma~\ref{lem7} by setting $\alpha=1$.
\end{IEEEproof}

\begin{IEEEproof}[Proof of Theorem~\ref{thm:TrEandTrH}]
Assume there exists a pair of $\F_{r}$-linear codes $C_{1}$ and $C_{2}$ of
length $n$ over $\F_{q}$ such that $C_{1}^{\perp_{\TrE}} \subseteq C_{2}$ with
$\frac{|C_{2}|}{\left|C_{1}^{\perp_{\TrE}}\right|}=K$,  $d_{x}=\wt(C_{1}\setminus C_{2}^{\perp_{\TrE}} )$ and $ d_{z}=\wt(C_{2}\setminus C_{1}^{\perp_{\TrE}})$.

\begin{enumerate}[\text{Case} 1.]
\item If $q$ is odd, then by Lemma~\ref{lem7} $i)$, we have
$\alpha^{-1}\overline{C_{1}^{\perp_{\TrE}}} \subseteq
\left( C_{2}^{\perp_{\TrE}} \right)^{\perp_{\Tr\H}}$ with
\begin{align*}
K &=\frac{|C_{2}|}{\left|C_{1}^{\perp_{\TrE}}\right|}=
\frac{\left|\left( C_{2}^{\perp_{\TrE}} \right)^{\perp_{\Tr\H}}\right|}
{\left|\alpha^{-1}\overline{C_{1}^{\perp_{\TrE}}}\right|}\text{.}
\end{align*}

Since the codes $\alpha^{-1}\overline{C_{1}^{\perp_{\TrE}}}$ and  $C_{2}^{\perp_{\TrE}}$ are equivalent and 
it follows from (\ref{eq:MacW}) that  $C_2=(C_2^{\perp_{\TrE}})^{\perp_{\TrE}}$ and 
$(C_2^{\perp_{\TrE}})^{\perp_{\Tr\H}}$ share the same weight enumerator, we have
\begin{align*} 
&\wt\left( \left( C_{2}^{\perp_{\TrE}} \right)^{\perp_{\Tr\H}}   \setminus  \alpha^{-1}\overline{C_{1}^{\perp_{\TrE}}} \right)\\
&= \wt(C_{2}\setminus C_{1}^{\perp_{\TrE}})  =d_{z}.
\end{align*}

The code $C_1$ is equivalent to $ (\alpha^{-1}\overline{C_1^{\perp_{\TrE}}})^{\perp_{\TrE}}$  
which, by (\ref{eq:MacW}), shares the same weight enumerator with  $(\alpha^{-1}\overline{C_1^{\perp_{\TrE}}})^{\perp_{\Tr\H}}$.  Hence
\begin{align*}&\wt\left( \left( \alpha^{-1}\overline{C_{1}^{\perp_{\TrE}}}\right)^{\perp_{Tr_{q/r}H}} \setminus C_{2}^{\perp_{\TrE}} \right)\\
&= \wt(C_{1}\setminus C_{2}^{\perp_{\TrE}} )=d_{x}.
\end{align*}
The conclusion follows from~\cite[Th. 4.5]{ELS10}.

\item If $q$ is even, then the proof is similar to that of Case 1 with
$\alpha = 1$ and using Lemma~\ref{lem8} $i)$ instead of Lemma~\ref{lem7} $i)$.
\end{enumerate}

Conversely, assume that there exists a pair of $\F_{r}$-linear codes
$C_{1}$ and $C_{2}$ of length $n$ over $\F_{q}$ such that
$C_{1}^{\perp_{\Tr\H}} \subseteq C_{2}$ with
$\frac{|C_{2}|}{\left|C_{1}^{\perp_{\Tr\H}}\right|}=K$, $d_{x}=\wt(C_{1}\setminus C_{2}^{\perp_{\Tr\H}} )$ and $ d_{z}=\wt(C_{2}\setminus C_{1}^{\perp_{\Tr\H}})$.

\begin{enumerate}[\text{Case} 1.]
\item If $q$ is odd, then by Lemma~\ref{lem7} $ii)$, we have
$\alpha \overline{C_{1}^{\perp_{\Tr\H}}} \subseteq
\left( C_{2}^{\perp_{\Tr\H}} \right)^{\perp_{\TrE}}$ with
\begin{align*}
 K &=\frac{|C_{2}|}{\left|C_{1}^{\perp_{\Tr\H}}\right|}=
\frac{\left|\left( C_{2}^{\perp_{\Tr\H}} \right)^{\perp_{\TrE}}\right|}
{\left|\alpha\overline{C_{1}^{\perp_{\Tr\H}}}\right|}\text{.}
\end{align*}
Using similar observation as in Case  1 of the  necessary part, we have  

\[\wt\left( \left( C_{2}^{\perp_{\Tr\H}} \right)^{\perp_{\TrE}} \setminus \alpha \overline{C_{1}^{\perp_{\Tr\H}}}\right) =d_{z}\]
and 
\[ \wt\left( \left(\alpha \overline{C_{1}^{\perp_{\Tr\H}}} \right)^{\perp_{\TrE}} \setminus C_{2}^{\perp_{\Tr\H}} \right)=d_{x}.\]

The conclusion follows from Theorem~\ref{thm:AQCTrEu}.
\item If $q$ is even, then the proof is similar to that of Case 1 with
$\alpha = 1$ and using Lemma~\ref{lem8} $ii)$ instead of Lemma~\ref{lem7} $ii)$.
\end{enumerate}
\end{IEEEproof}

\section*{Appendix C: Proof of Theorem~\ref{thm:cyclic_main}}
We will need the following lemma in the proof.

\renewcommand{\thetheorem}{C.1}
\begin{lemma}\label{lem:quadratic}
Let $\F_{q}=\F_{r}(\omega)$ be a quadratic extension of $\F_{r}$.
Then the following statements hold:
\begin{enumerate}[$i)$]
\item Any $(n, r^{l})_{q}$-cyclic $\F_{r}$-linear code $C$ over $\F_{q}$
has two generators and can be written as $C:=\langle a(x) +\omega b(x), c(x)\rangle$,
where $a(x), b(x)$, and $c (x)$ are polynomials in $\F_{r}[x]$,
$b(x)$ and $c(x)$ are monic divisors of $x^n -1$ in $\F_{r}[x]$,
$c (x)$ divides $a(x)(x^n -1)/b(x)$ in $\F_{r}[x]$, and
$l = 2n - \deg (b(x))- \deg(c (x))$.
\item If $\langle a^\prime(x) +\omega b^\prime(x), c^\prime(x)\rangle$
is another representation of $C$ in the above sense, then $b^\prime(x) = b(x)$,
$c ^\prime(x) = c (x)$ and $a^\prime(x) \equiv a(x) ({\rm mod}\, c (x))$.
\end{enumerate}
\end{lemma}
\begin{IEEEproof}
To prove $i)$, let $C$ be an $(n, r^{l})_{q}$-cyclic $\F_{r}$-linear code
over $\F_{q}$. Define an $\F_{r}[x]$-module homomorphism
\begin{align*}
\varphi : &C \rightarrow \F_{r}[x]/\langle x^n-1 \rangle \text{ sending} \\
   & v(x) :=f_0(x)+\omega f_1(x) \mapsto f_1(x) \text{,}
\end{align*}
where $f_0(x)$ and $f_1(x)$ are polynomials in $\F_{r}[x]$.

The zero code is viewed as the one generated by $x^n-1$.
The kernel $\ker(\varphi)=\{ v(x) \in C : f_{1}(x) \equiv 0 \}
= \{ f_{0}(x) \in C\}$ and the image $\varphi(C) = \{ f_{1}(x) : v(x) \in C \}$
are linear cyclic codes over $\F_{r}$. Hence, there exist unique, monic
generators $c(x)$ and $b(x)$ of minimal degree, respectively, such that
$\ker(\varphi) = \langle c(x) \rangle$ and $\varphi(C)=\langle b(x) \rangle$.

Note that for all $a(x) \in \F_{r}[x]$,
\begin{align}
 \dimr(C) &= \dimr(\langle c(x) \rangle) + \dimr(\langle b(x) \rangle) \notag \\
          &= \dimr(\langle c(x), \omega b(x) \rangle)\label{eq:second} \\
          & \leq \dimr(\langle c(x), a(x)+ \omega b(x) \rangle)\label{eq:third} \text{.}
\end{align}
The first equation is clear from the definition of $\varphi$. To justify (\ref{eq:second}), 
let $0 \leq i \leq n - \deg(c(x))-1$ and $ 0 \leq j\leq n-\deg(b(x))-1$. The sets $\{x^{i} c(x)\}$ and $\{x^{j}b(x)\}$ serve, 
respectively, as bases for $\langle c(x) \rangle$ and $\langle b(x) \rangle$. Since the set 
$\{x^{i} c(x), x^{j} \omega b(x)\}$ is $\F_{r}$-linearly independent,
\begin{align*}
	\dimr(\langle c(x), \omega b(x) \rangle) & \geq 2n-\deg(b(x))-\deg(c(x))\\
	                                         & =\dimr(\langle c(x) \rangle) + \dimr(\langle b(x) \rangle)\text{.}
\end{align*}
The other direction is clear.

To establish (\ref{eq:third}), notice that the set $\{x^{i} c(x), x^{j} (a(x) + \omega b(x))\} 
\subseteq \langle c(x), (a(x)+\omega b(x)) \rangle$ is again $\F_{r}$-linearly independent.

If $a(x)+\omega b(x)$ is a preimage of $b(x)$ under $\varphi$, then $\langle c(x), (a(x)+\omega b(x)) \rangle \subseteq C$. 
By (\ref{eq:third}), we conclude that $\langle c(x), (a(x)+\omega b(x)) \rangle = C$ and that $C$ has the claimed $\F_{r}$-dimension.

Clearly, $c(x)$ and $b(x)$ divide  $x^n-1$ in $\F_{r}[x]$.
Let $g(x):=(x^n-1)/b(x)$. Since
\begin{equation*}
a(x)g(x) \equiv (a(x)+\omega b(x)) {g(x)} \in \ker(\varphi)\text{,}
\end{equation*}
$c(x)$ divides $a(x)g(x)$ in $\F_{r}[x]$.

To prove $ii)$, assume $C:=\langle a^\prime(x) +\omega b^\prime(x), c^\prime(x)\rangle$.
Then $b(x)|b^\prime(x)$, $c(x)|c^\prime(x)$ and
\begin{align*}
\deg(b^\prime(x)c^\prime(x)) &= \deg(b^\prime(x))+\deg(c^\prime(x)) \\
                             &=2n-l = \deg(b(x))+\deg(c(x)) \\
                             & =\deg(b(x)c(x))\text{.}
\end{align*}
Hence, we have $b(x)=b^\prime(x)$ and $c(x)=c^\prime(x)$.

Since $a^\prime(x)+\omega b(x) \in \langle a(x)+\omega b(x), c(x)\rangle$, there
exist polynomials $s(x),t(x)\in \F_{r}[x]$ such that
\begin{equation*}
a^\prime(x)+\omega b(x)=s(x)(a(x)+\omega b(x))+t(x)c(x).
\end{equation*}
Without loss of generality, $\deg(s(x)) < n-\deg(b(x))$ and $\deg(t(x)) < n-\deg(c(x))$ can be assumed. 
By comparing the coefficients on both sides of the equation, $s(x) = 1$.
Therefore, $ a^\prime(x) = a(x)+t(x)c(x)$, making $ a^\prime(x) \equiv a(x)\,({\rm mod}\, c(x))$.
\end{IEEEproof}

With the lemma established, the theorem can now be settled.
\begin{IEEEproof}[Proof of Theorem~\ref{thm:cyclic_main}] 
We prove by induction on $m$. If $m=2$, the statement follows from
Lemma~\ref{lem:quadratic}. Assume that the theorem holds for $m-1$.
Let $C$ be an $(n, r^{l})_{q}$-cyclic $\F_{r}$-linear code  over $\F_{q}$ and
let $\F_{s}$ be the field extension of $\F_{r}$ of degree $m-1$ such
that $\F_{s}=\F_{r}(\alpha)$.

Let $\phi: C \rightarrow \F_{s}[x]/\langle x^n-1\rangle$ be an
$\F_{r}[x]$-module homomorphism defined by
\begin{align*}
& f_0(x)+\omega f_1(x)+\omega^2 f_2(x)+\dots+ \omega^{m-1} f_{m-1}(x) \\
& \mapsto f_1(x)+\alpha f_2(x)+\dots+ \alpha^{m-2} f_{m-1}(x)\text{.}
\end{align*}
The kernel $\ker(\phi)$ and the image $\phi(C)$ are a linear cyclic
code over $\F_{r}$ and a cyclic $\F_{r}$-linear code over $\F_{s}$,
respectively.

Let $a_{m-1,0}(x)$ be the unique monic generator of $\ker(\phi)$
of minimal degree. By the induction hypothesis, $\phi(C)$ has $m-1$
generators, say
\begin{align*}
\phi(C)=\langle & a_{0,1}(x) + \alpha a_{0,2}(x) + \ldots + \alpha^{m-2}a_{0,m-1}(x), \\
                & a_{1,1}(x) + \alpha a_{1,2}(x) + \ldots + \alpha^{m-3}a_{1,m-2}(x), \\
                & ~~~ \vdots ~~~~~~~~~~~      \vdots ~~~~~~~~~~~~~\iddots \\
                & a_{m-3,1}(x) + \alpha a_{m-3,2}(x),\\
                & a_{m-2,1}(x)\rangle,  \\
\end{align*}
satisfying properties $i)$ to $iv)$. Therefore, $C$ is an $\F_{r}[x]$-module
generated by
\begin{align*}
\langle & a_{0,0}(x) + \omega a_{0,1}(x) + \ldots + \omega^{m-1}a_{0,m-1}(x), \\
        & a_{1,0}(x) + \omega a_{1,1}(x) + \ldots + \omega^{m-2}a_{1,m-2}(x), \\
        & ~~ \vdots ~~~~~~~~~~~      \vdots ~~~~~~~~~~~~~\iddots \\
        &a_{m-2,0}(x) + \omega a_{m-2,1}(x),\\
        &a_{m-1,0}(x) \rangle,  \\
\end{align*}
where $ a_{i,0}(x) + \omega a_{i,1}(x) + \ldots + \omega^{m-1-i}a_{i,m-1-i}(x)$
is an inverse image of $a_{i,1}(x)+\alpha a_{i,2}(x)+\ldots+ \alpha^{m-2-i}a_{i,m-1-i}(x)$
for all $0\leq i\leq m-2$. Clearly, property $i)$ holds.

Using a similar reasoning to the proof of Lemma~\ref{lem:quadratic}, by the inductive hypothesis we obtain the fact that
$a_{m-2,0}(x)(x^n-1)/a_{m-2,1}(x) $ is divisible by $a_{m-1,0}(x)$. Hence, property $ii)$ follows.

Since
\begin{equation*}
\dimr(\ker(\phi))=n-\deg(a_{m-1,0}(x))
\end{equation*}
and
\begin{equation*}
\dimr(\phi(C))=(m-1)n-\sum\limits_{i=0}^{m-2}\deg(a_{i,m-1-i}(x)) \text{,}
\end{equation*}
we have
\begin{align*}
l &=\dimr(\ker(\phi))+\dimr( \phi(C)) \\
  &=mn-\sum\limits_{i=0}^{m-1}\deg(a_{i,m-1-i}(x)) \text{.}
\end{align*}
This proves property $iii)$.

The uniqueness stated in property $iv)$ can be obtained from an argument similar
to the one used in the proof of $b)$ in Lemma~\ref{lem:quadratic}.
\end{IEEEproof}

\section*{Appendix D:  Farkas Certificate of Infeasibility}
Referring to~\eqref{eq:Monetwo}, the tuple 
\begin{equation*}
(n, q,k,k',d_x,d_z)=(6,2,2,1,3,2)
\end{equation*}
has a vector $\mathbf{r}$ with $\mathbf{r}^\intercal=\left(1,0,0,0,1,0,0,\ldots,0\right)$ and the 
matrices $M_1$ and $M_2$ below.
\begin{align*}
M_1 &=
\left(
\begin{smallmatrix}
1 & 0 & 0 & 0 & 0 & 0 & 0 & 0 & 0 & 0 & 0 & 0 & 0 & 0 \\
0 & 1 & 0 & 0 & 0 & 0 & 0 & 0 & 0 & 0 & 0 & 0 & 0 & 0 \\
6 & 4 & 2 & 0 & -2 & -4 & -6 & 0 & 0 & 0 & 0 & 0 & 0 & 0 \\
15 & 5 & -1 & -3 & -1 & 5 & 15 & 0 & 0 & 0 & 0 & 0 & 0 & 0 \\
0 & 0 & 0 & 0 & 0 & 0 & 0 & 1 & 0 & 0 & 0 & 0 & 0 & 0 \\
0 & 0 & 0 & 0 & 0 & 0 & 0 & 0 & 1 & 0 & 0 & 0 & 0 & 0 \\
0 & 0 & 0 & 0 & 0 & 0 & 0 & 0 & 0 & 1 & 0 & 0 & 0 & 0 \\
0 & 0 & 0 & 0 & 0 & 0 & 0 & 6 & 4 & 2 & 0 & -2 & -4 & -6 \\
-1 & -1 & -1 & -1 & -1 & -1 & -1 & 32 & 0 & 0 & 0 & 0 & 0 & 0 \\
-6 & -4 & -2 & 0 & 2 & 4 & 6 & 0 & 32 & 0 & 0 & 0 & 0 & 0 \\
-15 & -5 & 1 & 3 & 1 & -5 & -15 & 0 & 0 & 32 & 0 & 0 & 0 & 0 \\
8 & 0 & 0 & 0 & 0 & 0 & 0 & -1 & -1 & -1 & -1 & -1 & -1 & -1 \\
0 & 8 & 0 & 0 & 0 & 0 & 0 & -6 & -4 & -2 & 0 & 2 & 4 & 6
\end{smallmatrix}
\right) \text{,}\\
M_2 &=
\left(
\begin{smallmatrix}
20 & 0 & -4 & 0 & 4 & 0 & -20 & 0 & 0 & 0 & 0 & 0 & 0 & 0 \\
15 & -5 & -1 & 3 & -1 & -5 & 15 & 0 & 0 & 0 & 0 & 0 & 0 & 0 \\
6 & -4 & 2 & 0 & -2 & 4 & -6 & 0 & 0 & 0 & 0 & 0 & 0 & 0 \\
1 & -1 & 1 & -1 & 1 & -1 & 1 & 0 & 0 & 0 & 0 & 0 & 0 & 0 \\
0 & 0 & 0 & 0 & 0 & 0 & 0 & 15 & 5 & -1 & -3 & -1 & 5 & 15 \\
0 & 0 & 0 & 0 & 0 & 0 & 0 & 20 & 0 & -4 & 0 & 4 & 0 & -20 \\
0 & 0 & 0 & 0 & 0 & 0 & 0 & 15 & -5 & -1 & 3 & -1 & -5 & 15 \\
0 & 0 & 0 & 0 & 0 & 0 & 0 & 6 & -4 & 2 & 0 & -2 & 4 & -6 \\
0 & 0 & 0 & 0 & 0 & 0 & 0 & 1 & -1 & 1 & -1 & 1 & -1 & 1 \\
-20 & 0 & 4 & 0 & -4 & 0 & 20 & 0 & 0 & 0 & 32 & 0 & 0 & 0 \\
-15 & 5 & 1 & -3 & 1 & 5 & -15 & 0 & 0 & 0 & 0 & 32 & 0 & 0 \\
-6 & 4 & -2 & 0 & 2 & -4 & 6 & 0 & 0 & 0 & 0 & 0 & 32 & 0 \\
-1 & 1 & -1 & 1 & -1 & 1 & -1 & 0 & 0 & 0 & 0 & 0 & 0 & 32 \\
0 & 0 & 8 & 0 & 0 & 0 & 0 & -15 & -5 & 1 & 3 & 1 & -5 & -15 \\
0 & 0 & 0 & 8 & 0 & 0 & 0 & -20 & 0 & 4 & 0 & -4 & 0 & 20 \\
0 & 0 & 0 & 0 & 8 & 0 & 0 & -15 & 5 & 1 & -3 & 1 & 5 & -15 \\
0 & 0 & 0 & 0 & 0 & 8 & 0 & -6 & 4 & -2 & 0 & 2 & -4 & 6 \\
0 & 0 & 0 & 0 & 0 & 0 & 8 & -1 & 1 & -1 & 1 & -1 & 1 & -1
\end{smallmatrix}
\right) \text{.}
\end{align*}

Letting
\begin{align*}
\mathbf{s}_1^\intercal &=
\left(0,\tfrac{14}{3},-\tfrac{2}{3},-\tfrac{3}{4},4,-1,-1,
-\tfrac{1}{12},-\tfrac{1}{3},-1,-1,-1,-1\right) \text{,}\\
\mathbf{s}_2^\intercal &= \left(\tfrac{1}{8},0,\tfrac{1}{32},0,0,0,0,0,1,0,0,
\tfrac{11}{96},\tfrac{1}{12},0,\tfrac{1}{24},0,0,0\right) \text{,}
\end{align*}
one can see that 
\[
\mathbf{s}_1^\intercal M_1+ \mathbf{s}_2^\intercal M_2=
\left(0,0,0,0,0,0,0,0,-\tfrac{88}{3},-29,0,0,\ldots,0\right),
\]
$\mathbf{s}_1^\intercal \mathbf{r}=4$ and $\mathbf{s}_2\geq \mathbf{0}$, 
as required by our criterion.

\section*{Acknowledgements}
We gratefully acknowledge IBM for CPLEX, which is the LP/ILP solver in IBM ILOG CPLEX 
Optimization Studio 12.2 that IBM provides free of charge via the IBM Academic Initiative. 
The authors thank Markus Grassl for useful discussions.

\begin{IEEEbiographynophoto}{Martianus~Frederic~Ezerman} (M'10) 
grew up in East Java Indonesia. He received the B.A. degree in philosophy 
and the B.Sc. degree in mathematics in 2005 as well as the M.Sc. degree 
in mathematics in 2007, all from Ateneo de Manila University, Philippines. 
In 2011 he obtained the Ph.D. degree in mathematics from Nanyang Technological 
University, Singapore. After a year of postdoctoral fellowship at Laboratoire d'Information Quantique, 
Universit{\'e} Libre de Bruxelles, Belgium, he joined the Centre for Quantum Technologies 
at the National University of Singapore in October of 2012.

His main research interests are error-correcting codes, both classical and quantum, and 
related structures in algebra and combinatorics.
\end{IEEEbiographynophoto}

\begin{IEEEbiographynophoto}{Somphong Jitman}
received the B.Sc. degree from Prince of Songkla University, Thailand,
in 2005 and the M.Sc. and Ph.D. degrees in mathematics from Chulalongkorn University,
Thailand, in 2007 and 2011, respectively. He was a Research Fellow with the Division of
Mathematical Sciences, School of Physical and Mathematical Sciences, at Nanyang Technological 
University in Singapore for two years prior to returning to Thailand in May of 2013 to become 
a Lecturer with the Department of Mathematics, Faculty of Science, Silpakorn University.

His research interests cover classical error-correcting codes,
codes over rings as well as quantum error-correcting codes.
\end{IEEEbiographynophoto}

\begin{IEEEbiographynophoto}{San~Ling}
received the B.A. degree in mathematics from the University of Cambridge 
in 1985 and the Ph.D. degree in mathematics from the University of California, 
Berkeley in 1990. Since April 2005, he has been a Professor with the Division of Mathematical 
Sciences, School of Physical and Mathematical Sciences, in Nanyang 
Technological University, Singapore. Prior to that, he was with the 
Department of Mathematics, National University of Singapore. 

His research fields include arithmetic of modular curves and 
applications of number theory to combinatorial designs, coding theory, 
cryptography and sequences.
\end{IEEEbiographynophoto}
\pagebreak
\begin{IEEEbiographynophoto}{Dmitrii V. Pasechnik}
received the M.Sc. degree in computing from Moscow Institute for Steel and Alloys in 1988 and
the Ph.D. degree in mathematics from the University of Western Australia in 1996.
Prior to joining  the Division of Mathematical Sciences, School of Physical and Mathematical Sciences, 
in Nanyang Technological University, Singapore in January 2006 he held a number of postdoctoral
positions in the Netherlands and Germany.

His research interests include combinatorics, optimisation, 
computational algebra and geometry, and complexity of algorithms. 
\end{IEEEbiographynophoto}

\end{document}